\documentclass{aastex631}

\newcommand\tde{AT2018hyz}

\usepackage{amsmath,amstext}
\usepackage[T1]{fontenc}
\usepackage[figure,figure*]{hypcap}
\usepackage{graphicx}
\usepackage{subfigure}
\usepackage{tabularx}
\usepackage{enumitem}
\usepackage{appendix}

\shorttitle{Late-time Radio in AT2018hyz}
\shortauthors{Cendes et al.}

\graphicspath{{./}{figures/}}

\begin{document}

\title[AT2018hyz]{A Mildly Relativistic Outflow Launched Two Years after Disruption in the Tidal Disruption Event AT2018hyz}

\correspondingauthor{Yvette Cendes}
\email{yvette.cendes@cfa.harvard.edu}

\author[0000-0001-7007-6295]{Y. Cendes}
\affiliation{Center for Astrophysics | Harvard \& Smithsonian, Cambridge, MA 02138, USA}

\author[0000-0002-9392-9681]{E. Berger}
\affiliation{Center for Astrophysics | Harvard \& Smithsonian, Cambridge, MA 02138, USA}

\author[0000-0002-8297-2473]{K. D. Alexander}
\affiliation{Center for Interdisciplinary Exploration and Research in Astrophysics (CIERA) and Department of Physics and Astronomy, Northwestern University, Evanston, IL 60208, USA}

\author[0000-0001-6395-6702]{S. Gomez}
\affiliation{Space Telescope Science Institute, 3700 San Martin Dr, Baltimore, MD 21218, USA}

\author[0000-0003-2349-101X]{A. Hajela}
\affiliation{Center for Interdisciplinary Exploration and Research in Astrophysics (CIERA) and Department of Physics and Astronomy, Northwestern University, Evanston, IL 60208, USA}

\author[0000-0002-7706-5668]{R. Chornock}
\affiliation{Department of Astronomy, University of California, Berkeley, CA 94720-3411, USA}

\author[0000-0003-1792-2338]{T. Laskar}
\affiliation{Department of Astrophysics/IMAPP, Radboud University, PO Box 9010, 6500 GL, The Netherlands}

\author[0000-0003-4768-7586]{R. Margutti}
\affiliation{Department of Astronomy, University of California, Berkeley, CA 94720-3411, USA}

\author[0000-0002-4670-7509]{B. Metzger}
\affiliation{Department of Physics and Columbia Astrophysics Laboratory, Columbia University, Pupin Hall, New York, NY 10027, USA}
\affiliation{Center for Computational Astrophysics, Flatiron Institute, 162 5th Ave, New York, NY 10010, USA}

\author[0000-0002-0592-4152]{M. F. Bietenholz}
\affiliation{Department of Physics and Astronomy, York University, 4700 Keele St., Toronto, M3J~1P3, Ontario, Canada}

\author[0000-0001-6415-0903]{D. Brethauer}
\affiliation{Department of Astronomy, University of California, Berkeley, CA 94720-3411, USA}

\author[0000-0002-7721-8660]{M. H. Wieringa}
\affiliation{CSIRO Space and Astronomy,PO Box 76,Epping NSW 1710,Australia}

\begin{abstract}
We present late-time radio/millimeter (as well as optical/UV and X-ray) detections of the tidal disruption event (TDE) AT2018hyz, 
spanning $970 - 1300$ d  after optical discovery.  In conjunction with earlier deeper limits, including at $\approx 700$ d, our observations reveal rapidly rising emission at $0.8-240$ GHz, steeper than $F_\nu\propto t^5$ relative to the time of optical discovery. Such a steep rise cannot be explained in any reasonable scenario of an outflow launched at the time of disruption (e.g., off-axis jet, sudden increase in the ambient density), and instead points to a delayed launch. Our multi-frequency data allow us to directly determine the radius and energy of the radio-emitting outflow, showing that it was launched $\approx 750$ d after optical discovery.  The outflow velocity is mildly relativistic, with $\beta\approx 0.25$ and $\approx 0.6$ for a spherical and a $10^\circ$ jet geometry, respectively, and the minimum kinetic energy is $E_K\approx 5.8\times 10^{49}$ and $\approx 6.3\times 10^{49}$ erg, respectively.  This is the first definitive evidence for the production of a delayed mildly-relativistic outflow in a TDE; a comparison to the recently-published radio light curve of ASASSN-15oi suggests that the final re-brightening observed in that event (at a single frequency and time) may be due to a similar outflow with a comparable velocity and energy.  Finally, we note that the energy and velocity of the delayed outflow in \tde\ are intermediate between those of past non-relativistic TDEs (e.g., ASASSN-14li, AT2019dsg) and the relativistic TDE Sw\,J1644+57. We suggest that such delayed outflows may be common in TDEs.
\end{abstract}


\section{Introduction} 
\label{sec:intro}

A tidal disruption event (TDE) occurs when a star wanders sufficiently close to a supermassive black hole (SMBH) to be torn apart by tidal forces, leading to the eventual formation of a transitory accretion flow \citep{Rees1988,Komossa2015}. Optical/UV and X-ray observations of TDEs are generally thought to track the mass fallback and accretion (e.g., \citealt{Stone2013,Guillochon2013}).  Radio observations, on the other hand, can reveal and characterize outflows from TDEs \citep{Alexander2020}, including the presence of relativistic jets \citep{Zauderer2011,Giannios11,DeColle2012}.

To date, rapid follow-up of TDEs, within days to weeks after discovery, have led to the detection of a few events.  These included most prominently the TDE Swift J1644+57 (Sw\,J1644+57), whose radio and mm emission were powered by a relativistic outflow with an energy of $\sim 10^{52}$ erg and an initial Lorentz factor of $\Gamma\sim 10$ \citep{Zauderer2011,Metzger2012,p1,p2,Eftekhari2018,Cendes2021}. Other events, such as ASASSN-14li and AT2019dsg, have instead exhibited evidence for non-relativistic outflows, with $E_K\sim 10^{48}-10^{49}$ erg and $\beta\approx 0.05-0.1$ \citep[e.g. ][]{Alexander2020,Alexander2016,Cendes2021b,Stein2020}.

Recently, two TDEs have been reported to show radio emission with a delay relative to the time of optical discovery.  ASASSN-15oi was first detected $\approx 180$ days after optical discovery with a luminosity that exceeded earlier radio limits (at 8, 23, and 90 days) by a factor of $\approx 20$ \citep{Horesh2021}. The radio emission subsequently declined until about 550 days, and then exhibited a second rapid rise with a detection at 1400 days with an even higher luminosity than the first peak; see Figure~\ref{fig:lumin-tde}.  iPTF16fnl was first detected $\approx 150$ days after optical discovery, with a luminosity about a factor of 8 times larger than earlier limits (extending to 63 days) and appeared to slowly brighten to about 417 days \citep{Horesh2021b}.  The initial abrupt rise in ASASSN-15oi seems distinct from the radio light curve of AT2019dsg, although both reach their peak radio luminosity on a similar timescale and at a similar level.  The gradual rise and much lower peak luminosity of iPTF16fnl ($\approx 10^{37}$ erg s$^{-1}$), on the other hand, may indicate that it is simply a less energetic example of typical radio-emitting TDEs.

Delayed radio emission from TDEs has been speculated to result from several possible effects.  First, it may be due to a decelerating off-axis relativistic jet launched at the time of disruption (e.g., \citealt{Giannios11,Mimica2015,Generozov2017}). Second, it may be due to an initial propagation of the outflow in a low density medium, followed by interaction with a significant density enhancement (e.g., \citealt{NakarGranot2007}). Finally, it may be due to a delayed launch of the outflow compared to the time of disruption and optical/UV emission, for instance as might result from a state transition in the accretion disk (e.g., \citealt{Tchekhovskoy2014}) or the delayed accumulation of magnetic flux onto the black hole (e.g., \citealt{Kelley2014}).  These scenarios can in principle be distinguished through a combination of detailed temporal and spectral information, that can be used to infer outflow properties such as radius, velocity, and energy, as well as the ambient density and its radial structure.  In the case of ASASSN-15oi it has been speculated that a delayed outflow may best explain the radio data \citep{Horesh2021}.

Against this backdrop, here we report the detection of rapidly rising radio emission from \tde\ ($z=0.0457$) starting about 970 days after optical discovery, with a factor of 30 increase in luminosity compared to upper limits at about 700 days.  Our extensive multi-frequency data, spanning 0.8 to 240 GHz, as well as optical/UV and X-rays, allow us to characterize the rising phase of emission in detail for the first time and hence to distinguish between possible scenarios for such a late rise.  We find that the radio emission requires an energetic ($E_K\approx 10^{50}$ erg) mildly relativistic ($\beta\approx 0.2-0.6$) outflow launched with a significant delay of about $700-750$ days after optical discovery. An off-axis relativistic jet launched near the time of optical discovery, or a large density enhancement can be ruled out.

The paper is structured as follows.  In \S\ref{sec:obs} we describe our new late-time radio, mm, UV/optical, and X-ray observations, and in \S\ref{sec:lumin} we contrast the radio emission from \tde\ with those of previous TDEs.  In \S\ref{sec:modeling} we model the radio spectral energy distribution and carry out an equipartition analysis to derive the physical properties of the outflow and environment.  In \S\ref{sec:params} we describe the results for a spherical and a collimated outflow geometry.  We discuss the implications of a delayed mildly-relativistic outflow in \S\ref{sec:disc} and summarize our findings in \S\ref{sec:conc}.

\section{Observations}
\label{sec:obs}

\subsection{Radio Observations}
\label{sec:obs-radio}

\tde\ was observed in targeted observations with AMI-LA 32 days after optical discovery leading to an upper limit of $F_\nu(15.5\,{\rm GHz}) \lesssim 85$ $\mu$Jy \citep{Horesh2018}, and with ALMA at 45 and 66 days to limits of $F_\nu(100\,{\rm GHz})\lesssim 38$ and $48 \mu$Jy, respectively \citep{Gomez2020}. The location of \tde\ was also covered in the ASKAP Variables and Slow Transients Survey \citep[VAST; ][]{Murphy2021} at 697 days with a limit of $F_\nu(0.9\,{\rm GHz})\lesssim 0.9$ mJy, and in the VLA Sky Survey epoch 2.1 \citep[VLASS; ][]{Lacy2020} at 705 days with a limit of $F_\nu(3\,{\rm GHz})\lesssim 0.45$ mJy. We determine the flux density limits in the survey images with the {\tt imtool fitsrc} command within the {\tt pwkit} package\footnote{https://github.com/pkgw/pwkit} \citep{Williams2017}.

As part of a broader study of late-time radio emission from TDEs, we observed \tde\ with the Karl G.~Jansky Very Large Array (VLA) at 972 days in C-band (Program ID 21A-303, PI: Hajela) and detected a source with a $F_\nu(5\,{\rm GHz})=1.39\pm 0.02$ mJy and $F_\nu(7\,{\rm GHz})=1.00\pm 0.02$ mJy (see Table~\ref{tab:obs}). Following this initial detection we obtained multi-frequency observations spanning L- to K-band ($\approx 1-23$ GHz; Programs 21B-357, 21B-360 and 22A-458, PI: Cendes), which resulted in detections across the full frequency range.  For all observations we used the primary calibrator 3C147, and the secondary calibrator J1024-0052.  We processed the VLA data using standard data reduction procedures in the Common Astronomy Software Application package (CASA; \citealt{McMullin2007}), using {\tt tclean} on the calibrated measurement set available in the NRAO archive.  The observations and resulting flux density measurements are summarized in Table~\ref{tab:obs}.  Additionally, we use data collected by the commensal VLA Low-band Ionosphere and Transient Experiment \citep[VLITE; ][]{Clarke2016} at 350 MHz during our multi-frequency observations (Table~\ref{tab:obs}).

We also obtained observations with the MeerKAT radio telescope in UHF and L-band ($0.8- 2$ GHz) on 2022 April 17 (1282 days; DDT-20220414-YC-01, PI: Cendes) and with the Australian Telescope Compact Array (ATCA) on 2022 May 1 (1296 days; Program C3472; PI: Cendes) at $2-20$ GHz. For ATCA we reduced the data using the {\tt MIRIAD} package. The calibrator 1934-638 was used to calibrate absolute flux-density and band-pass, while the calibrator 1038+064 was used to correct short term gain and phase changes. The {\tt invert}, {\tt mfclean} and {\tt restor} tasks were used to make deconvolved wideband, natural weighted images in each frequency band. For MeerKAT, we used the flux calibrator 0408-6545 and the gain calibrator 3C237, and used the calibrated images obtained via the SARAO Science Data Processor (SDP)\footnote{https://skaafrica.atlassian.net/wiki/spaces/ESDKB/pages/338723406/}.  We confirmed via the secondary SDP products that the source fluxes in the MeerKAT images were $\sim90\%$ of the sources overlapping with the NRAO VLA Sky Survey \citep[NVSS; ][]{Condon1998}, and that frequency slices show a steady increase of flux within the MeerKAT frequency range (with a spectral index $\alpha \approx -1$).

\subsection{Millimeter Observations}

Following the initial VLA radio detection we also observed \tde\ with the Atacama Large Millimeter/submillimeter Array (ALMA) at 1141, 1201, and 1253 days (Project 2021.1.01210.T, PI: Alexander). These observations roughly coincide with the three multi-frequency VLA observations. The first and third ALMA observations were in band 3 (mean frequency of 97.5 GHz) and band 6 (240 GHz); the second epoch was in band 3 only.  For the ALMA observations, we used the standard NRAO pipeline (version: 2021.2.0.128) in CASA (version: 6.2.1.7) to calibrate and image the data. We detect \tde\ in all observations with a rising flux density (Table \ref{tab:obs}).  We note that in the first observation, the flux density is at least 30 times higher than the early upper limits.

\startlongtable
\begin{deluxetable*}{lccllll}
\label{tab:obs}
\tablecolumns{7}
\tablecaption{Radio and Millimeter Observations of \tde}
\tablehead{
	\colhead{Date} &
	\colhead{Observatory} &
	\colhead{Project} &
	\colhead{$\delta t$\,$^a$} &
	\colhead{$\nu$} &
	\colhead{$F_\nu$\,$^b$} &
	\colhead{Source}
	\\
    \colhead{} &
    \colhead{} &
    \colhead{} &
    \colhead{(d)} &
    \colhead{(GHz)} &
    \colhead{(mJy)} &
    \colhead{}
}
\startdata
2018 Nov 15 & AMI & & 32 & 15.5 & $<0.085$ & \citet{Horesh2018}\\
\hline 
2018 Nov 28 & ALMA & & 45 & 97.5 & $<0.038$ & \citet{Gomez2020}\\
\hline 
2018 Dec 19 & ALMA & & 66 & 97.5 & $<0.043$ & \citet{Gomez2020}\\
\hline
2020 Sep 9 & ASKAP & VAST& 697 & 0.89 & $<0.9$ & This Work\\
\hline
2020 Sep 17 & VLA & VLASS 2& 705 & 3 & $<0.45$ & This Work\\
\hline 
2021 Jun 11 & VLA & 21A-303& 972 & 5 & 1.388 $\pm$ 0.019 & This Work\\
& & &  & 7 & 1.000 $\pm$ 0.018\\
\hline
2021 July 22 & ASKAP & VAST & 1013 & 0.89 & 1.30 $\pm$ 0.03 & \citet{Horesh2022} \\
\hline
2021 Nov 12 &VLA & VLITE & 1126 & 0.34 & <4.4 & This Work\\
& & 21B-357 &  & 1.37 & 4.753$\pm$0.084\\
& &  &  & 1.62 & 4.896$\pm$0.072\\
& &  &  & 1.88 & 4.799$\pm$0.078\\
& &  &  & 2.5 & 4.330$\pm$0.012\\
& &  &  & 3.5 & 3.668$\pm$0.051\\
& &  &  & 5 &2.939$\pm$0.030\\
& &  &  &  7&2.327$\pm$0.027\\
& &  &  & 9 & 2.030$\pm$0.027\\
& &  &  & 11 & $1.704\pm$0.033\\
& &  &  &  14& 1.560$\pm$0.052\\
& &  &  &  17& 1.262$\pm$0.011\\
& &  &  &  20& 1.067$\pm$0.024\\
& &  &  &  23& 0.980$\pm$0.020\\
\hline
2021 Nov 26&ALMA & 2021.1.01210.T & 1141 &  97.5& 0.451$\pm$0.029 & This Work\\
& & &  &  240& 0.198$\pm$0.024\\
\hline 
2022 Jan 24 &VLA & VLITE & 1199 & 0.34 & <5.2 & This Work\\
 & & 21B-360&& 1.12 & 7.656$\pm$0.200 &\\
& &  &  & 1.37 & 8.416$\pm$0.119\\
& &  &  & 1.62 & 8.416$\pm$0.119\\
& &  &  & 1.88 & 8.134$\pm$0.087\\
& &  &  & 2.5 & 6.393$\pm$0.113\\
& &  &  & 3.5 & 5.425$\pm$0.061\\
& &  &  & 5 &4.381$\pm$0.036\\
& &  &  &  7&3.576$\pm$0.046\\
& &  &  & 9 & 3.110$\pm$0.052\\
& &  &  & 11 & 2.862$\pm$0.055\\
& &  &  &  14& 2.564$\pm$0.055\\
& &  &  &  17& 2.258$\pm$0.024\\
& &  &  &  20& 1.974$\pm$0.042\\
& &  &  &  23& 1.726$\pm$0.023\\
\hline
2022 Jan 26 &ALMA & 2021.1.01210.T & 1201 &  97.5& 0.769$\pm$0.023 & This Work\\
\hline
2022 Mar 17 &VLA & VLITE & 1251 & 0.34 & <4.0 & This Work\\
 & & 22A458&& 1.12 & 8.449$\pm$0.236\\
& &  &  & 1.37 & 8.740$\pm$0.093\\
& &  &  & 1.62 & 8.712$\pm$0.112\\
& &  &  & 1.88 & 8.649$\pm$0.109\\
& &  &  & 2.5 & 7.743$\pm$0.115\\
& &  &  & 3.5 & 6.658$\pm$0.146\\
& &  &  & 5 &4.807$\pm$0.209\\
& &  &  &  7&4.807$\pm$0.209\\
& &  &  & 9 & 4.249$\pm$0.124\\
& &  &  & 11 & 3.789$\pm$0.106\\
& &  &  &  14& 4.424$\pm$0.052\\
& &  &  &  17& 2.886$\pm$0.050\\
& &  &  &  20& 2.541$\pm$0.078\\
& &  &  &  23& 2.107$\pm$0.831\\
\hline
2022 Mar 19 &ALMA & 2021.1.01210.T & 1253 &  97.5& 1.264$\pm$0.018 & This Work\\
& & &  &  240& 0.642$\pm$0.021\\
\hline
2022 Apr 17 &MeerKAT & DDT-20220414-YC-01 & 1282 & 0.82 & 3.162$\pm$0.040 & This Work\\
& & &  &  1.3& 5.325$\pm$0.041\\
\hline
2022 May 1 &ATCA & C3472 & 1296 & 2.1 & 7.042$\pm$0.138 & This Work\\
& & &  &  5.5& 7.837$\pm$0.140\\
& & &  &  9.0& 6.153$\pm$0.137\\
& & &  &  17& 3.734$\pm$0.120\\
& & &  &  19& 3.424$\pm$0.123\\
\hline
\enddata
\tablecomments{$^a$ These values are measured relative to the time of optical discovery, 2018 Oct 14.\\ $^b$ Limits are $3\sigma$.}
\end{deluxetable*}

\subsection{X-ray Observations}
\label{sec:obs-xray}

We obtained a Director's Discretionary Time observation of \tde\ on 2022 March 19 ($\delta t=1253$ days) with ACIS-S onboard the \textit{Chandra} X-ray Observatory with an exposure time of 14.9 ks (Program 23708833, PI: Cendes). We processed the data with \texttt{chandra$\_$repro} within \texttt{CIAO} 4.14 using the latest calibration files. An X-ray source is detected with \texttt{wavdetect} at the position of AT2018hyz with a net count-rate of $(8.9\pm2.5)\times 10^{-4}\,\rm{c\,s^{-1}}$ ($0.5-8$ keV), with statistical confidence of $6\,\sigma$ (Gaussian equivalent).

We extracted a spectrum of the source with $\texttt{specextract}$ using a $1.5''$ radius source region and a source-free background region of $22''$ radius. We fit the spectrum with an absorbed simple power-law model ($\texttt{tbabs*ztbabs*pow}$ within $\texttt{Xspec}$). The Galactic neutral hydrogen column density in the direction of AT2018hyz is $N_{H,\rm MW}=2.67\times 10^{20}\,\rm{cm^{-2}}$ \citep{Kalberla05}, and we find no evidence for additional intrinsic absorption, to a $3\sigma$ upper limit of $N_{H,\rm int}\lesssim 2.8\times 10^{22}\,\rm{cm^{-2}}$. The photon index is $\Gamma=1.5\pm 0.7$ ($1\,\sigma$ confidence level), and the $0.3-10$ keV unabsorbed flux is $F_x=1.78_{-0.28}^{+0.35} \times 10^{-14}\,\rm{erg\,s^{-1}cm^{-2}}$.  The flux density at 1 keV is $4.03\pm1.20\times10^{-6}$ mJy.

The X-ray flux is a factor of about 2 lower than the flux measured with XRT in the first 86 days, $F_x=4.1_{-0.4}^{+0.6} \times 10^{-14}\,\rm{erg\,s^{-1}cm^{-2}}$ \citep{Gomez2020}.  This indicates that \tde\ has faded in X-rays over this time scale, and we discuss the implications of this in \S\ref{sec:cooling}.

\subsection{UV/Optical Observations}


We obtained UV observations of AT\,2018hyz from the UV/Optical Telescope (UVOT; \citealt{Roming05}) on board the Neil Gehrels \textit{Swift} observatory (\textit{Swift}; \citealt{Gehrels04}) on 2022 January 7 (1182 days post-discovery) to continue tracking the evolution of the UV light curve originally presented in \cite{Gomez2020}. We measure an AB magnitude of $m_{\rm UVW2}= 20.20\pm 0.08$ mag using a $5''$ aperture centered on the position of AT\,2018hyz using the HEAsoft {\tt uvotsource} function \citep{Nasa14}. We correct this magnitude for Galactic extinction using the \citet{Barbary16} implementation of the \citet{Cardelli89} extinction law ($0.28$ mag) and subtract the host contribution of $21.76$ mag \citep{Gomez2020} to obtain a final value of $m_{\rm UVW2}=20.14 \pm 0.08$ mag.

We compare this measurement to the {\tt MOSFiT} model of AT\,2018hyz presented in \citet{Gomez2020}, which fit all optical/UV data to about 300 days post discovery. 
The model predicts a UVW2 magnitude of $23.3\pm 0.2$ at the time of our new observation, a factor of 18 times dimmer than observed.  This points to excess emission in the UV band.

Similarly, we measure the late time $r$-band magnitude of AT\,2018hyz at $\sim 1030$ days post-discovery by downloading raw ZTF images from the NASA/IPAC Infrared Science Archive\footnote{\url{https://irsa.ipac.caltech.edu/Missions/ztf.html}} and combining all $r$-band images taken within $\pm 12$ days of MJD = 59501. We subtract a template archival pre-explosion ZTF image from the combined image using {\tt HOTPANTS} \citep{Becker15}, and perform PSF photometry on the residual image. For calibration, we estimate the zero-point by measuring the magnitudes of field stars and comparing to photometric AB magnitudes from the PS1/$3\pi$ catalog. Corrected for galactic extinction, we find $r = 21.83 \pm 0.36$ mag. At this phase, the {\tt MOSFiT} model predicts a magnitude of $r = 24.65\pm 0.30$, a factor of 13 times dimmer than measured. We do not detect emission in $g$-band to a $3\sigma$ limit of $21.76$ mag. This implies a late time color of $g-r\gtrsim -0.1$, redder than the latest measurements from \citet{Gomez2020} of $g -r = -0.58 \pm 0.23$.

\section{Radio Luminosity and Evolution}
\label{sec:lumin}

\begin{figure}
\begin{center}
    \includegraphics[width=.75\columnwidth]{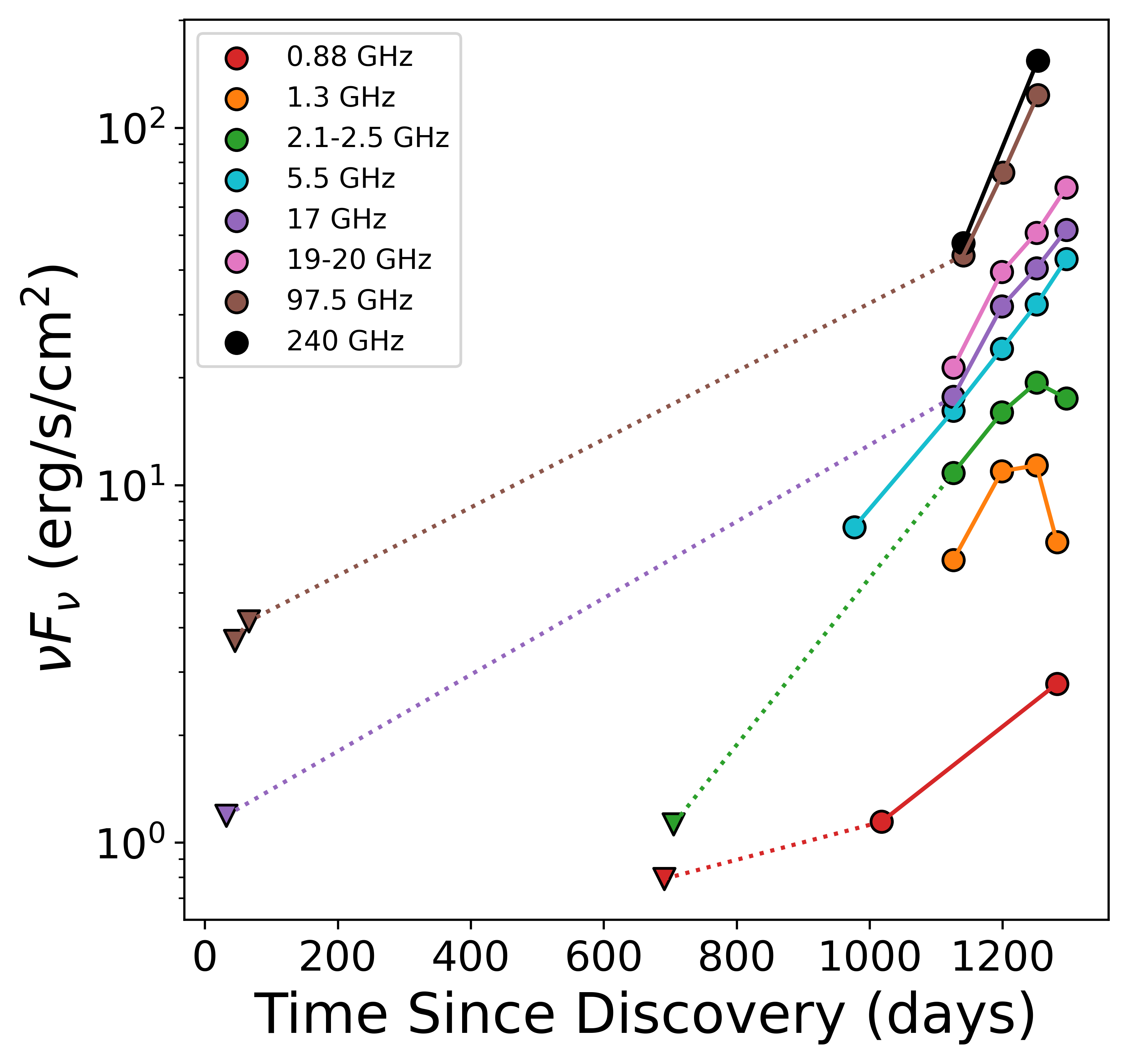}
    \end{center}
    \label{fig:bands}
    \caption{Luminosity light curve over time of AT2018hyz in several frequency bands, including early upper limits (triangles) and the late-time detections starting at about 970 days (circles).  While the source is rising in all frequencies during the first radio detections, we find the source has begun to fade in L-band (1.4 GHz, yellow) and S-band (3.0 GHz, green) after $\sim1250$ days.  In contrast, at higher frequencies such as C-band (5.5 GHz, light blue), X-band (9 GHz, dark blue), Ka-band (14 GHz, purple), K-band (19-20 GHz, pink), and in the millimeter band (97.5 GHz, brown, and 240 GHz, black) the source is still rising as roughly $F_\nu\propto t^5$ through 1300 days.  In the UHF-band (0.88 GHz, red) we see the source has risen in luminosity $\sim2.25\times$ from 1000 to 1280 days, but do not have enough sampling to establish whether it is decreasing.}
\end{figure}

The radio light curves of \tde\ at frequencies of $\approx 0.9-240$ GHz are shown in Figure~\ref{fig:bands}.  At all well-sampled frequencies we find a rapid rise at $\gtrsim 950$ days.  At S-band (3 GHz) the flux density rises by at least a factor of 16 from the non-detection at 705 days to a peak at about 1250 days. This corresponds to a steep power law rise ($F_\nu\propto t^\alpha$) with $\alpha\gtrsim 4.8$. Similarly, at C-band ($5-7$ GHz) we find a steady rise from about 1.4 mJy (972 days) to 7.8 mJy (1296 days) corresponding to $\alpha\approx 6$.  A similarly steep rise is observed up to 240 GHz. 
Such a steep rise occurring across a large spectral range is not expected in any model of delayed emission due to an off-axis viewing angle, a decelerating outflow, or a rapid increase in the ambient density (e.g., \citealt{Nakar2011}; see \S\ref{sec:params}).  Instead, the inferred steep power law rise indicates that the launch time of the outflow actually occurred much later than the time of optical discovery; for example, to achieve a power law rise of $t^3$, as expected for a decelerating outflow in a uniform density medium, requires a delay launch of $\sim 600$ days after optical discovery.  

We note that at frequencies of $\lesssim 3$ GHz, our latest observation indicates divergent behavior relative to the higher frequencies, with a pronounced decline in the flux density. For example, in L-band (1.4 GHz) we find a rapid decline from 8.7 to 5.3 mJy in the span of only 31 days (1251 to 1282 days).  This differential behavior is due to rapid evolution in the shape of the spectral energy distribution (see \S\ref{sec:equi}).

\begin{figure}
\begin{center}
    \includegraphics[width=.75\columnwidth]{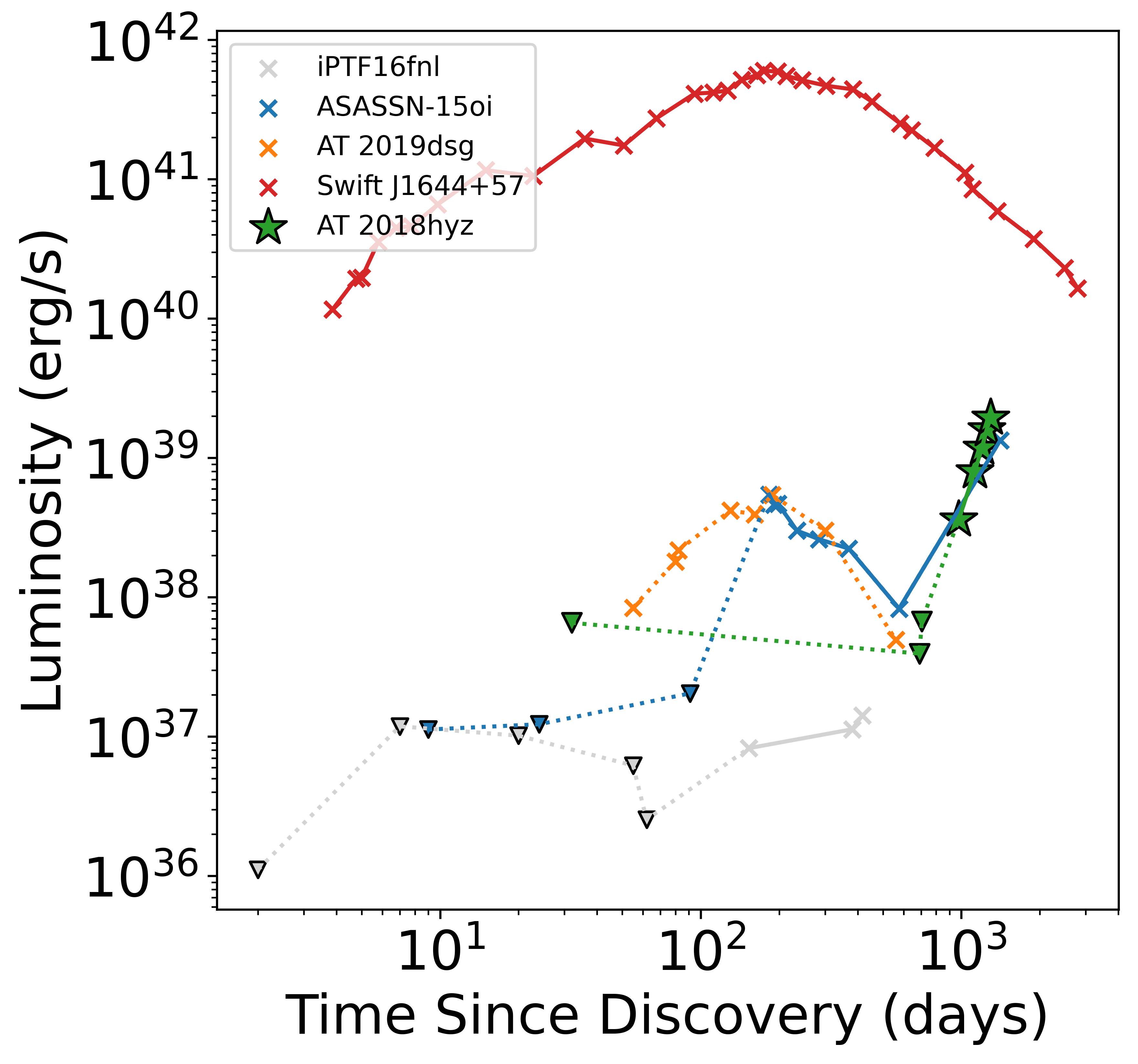}
    \end{center}
    \label{fig:lumin-tde}
    \caption{Luminosity light curve of AT2018hyz, including early upper limits (green triangles; 0.9, 3, and 15 GHz) and the late-time detections starting at about 970 days (green stars; 5 GHz).  Also shown for comparison are the light curves of the relativistic TDE Sw\,J1644+57 at (6.7 GHz; red; \citealt{p1,p2,Eftekhari2018,Cendes2021}), the non-relativistic event AT2019dsg (6.7 GHz; orange; \citealt{Cendes2021b}), and two events with apparent late-rising radio emission: ASASSN-15oi (6-7 GHz; blue; \citealt{Horesh2021}) and iPTF16fnl (15.5 GHz; grey; \citealt{Horesh2021b}).}
\end{figure}

In Figure~\ref{fig:lumin-tde} we show the radio light curve of \tde\ in the context of previous radio-emitting TDEs. The radio luminosity of \tde\ rapidly increases from $\lesssim 7\times 10^{37}$ erg s$^{-1}$ at $\approx 700$ days to $\approx 2\times 10^{39}$ erg s$^{-1}$ at $\approx 1300$ days, making it more luminous than any previous non-relativistic TDE. The rapid rise in \tde\ is even steeper than the second rising phase of ASASSN-15oi \citep[see Figure~\ref{fig:lumin-tde}; ][]{Horesh2021}, although the light curve of the latter contains only 2 data points (at 550 and 1400 days) and its actual rise may be steeper and comparable to \tde.  We also note that due to the wide gap in radio coverage of \tde\ between about 80 and 700 days, as well as the relatively shallower early radio limits compared to ASASSN-15oi, it is possible to ``hide'' an initial bump in the light curve as seen in ASASSN-15oi at $\approx 180-550$ days (Figure~\ref{fig:lumin-tde}); indeed, it is even possible that \tde\ had early radio emission comparable to that of AT2019dsg (\citealt{Cendes2021b}; Figure~\ref{fig:lumin-tde}), which had a nearly identical radio peak luminosity and timescale to ASASSN-15oi, but a more gradual and earlier rise.

Finally, we note that the radio emission from \tde\ is still about a factor of 20 times dimmer than that of Sw\,J1644+57 at a comparable timescale (1300 days) and about 80 times dimmer than its peak luminosity (Figure~\ref{fig:lumin-tde}). Since the powerful outflow in Sw\,J1644+57, with an energy of $\approx 10^{52}$ erg became non-relativistic at $\approx 700$ days \citep{Eftekhari2018}, this again argues against an off-axis jet interpretation for the less luminous (and hence less energetic) radio emission in \tde; namely, in such a scenario the radio emission would have peaked significantly earlier and with a much higher luminosity.

In the subsequent sections we model the radio SEDs to extract the physical properties of the outflow and ambient medium, as well as their time evolution, and show that these confirm our basic arguments for a delayed outflow.

\begin{figure}
\begin{center}
\includegraphics[width=.45\columnwidth]{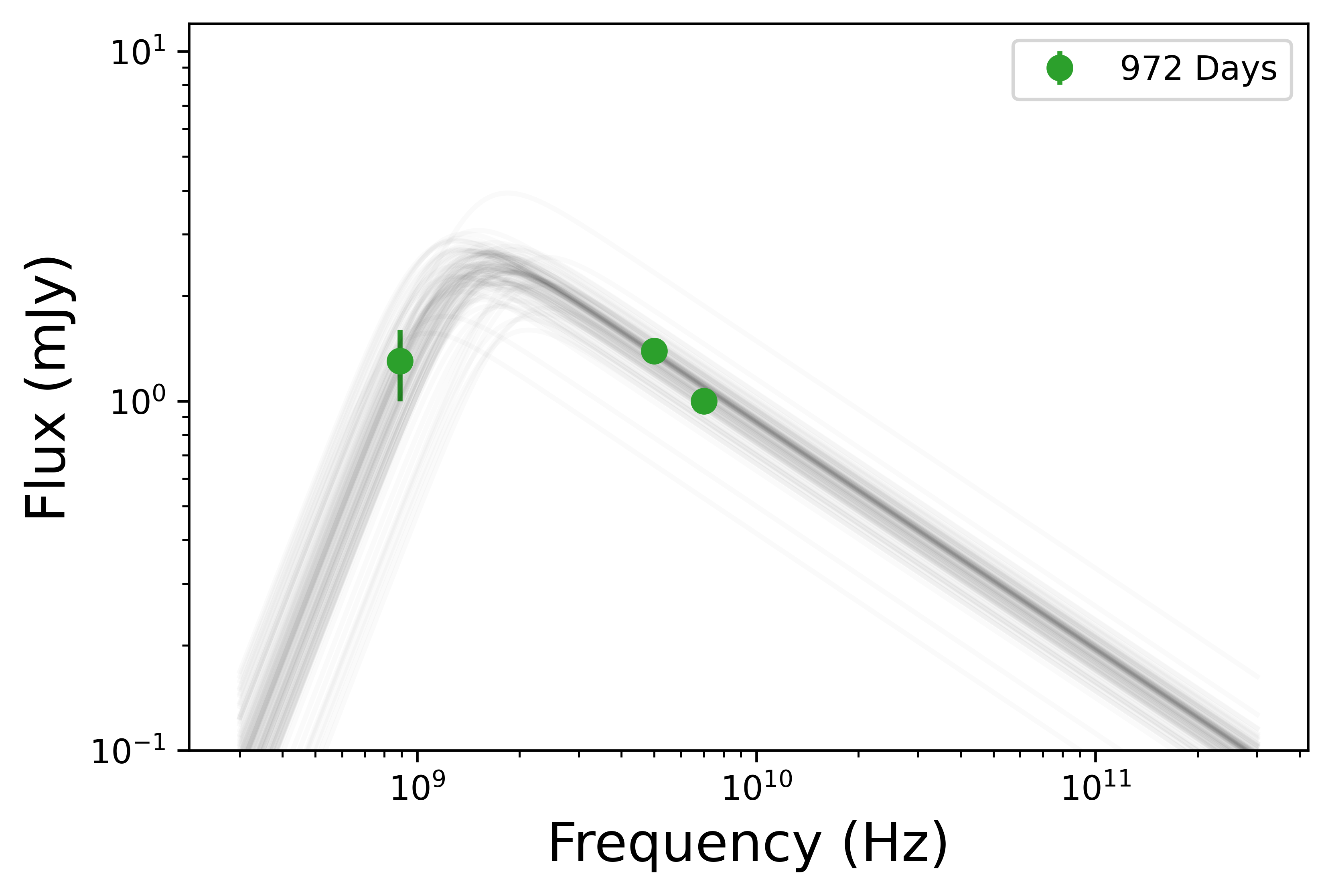}
\includegraphics[width=.45\columnwidth]{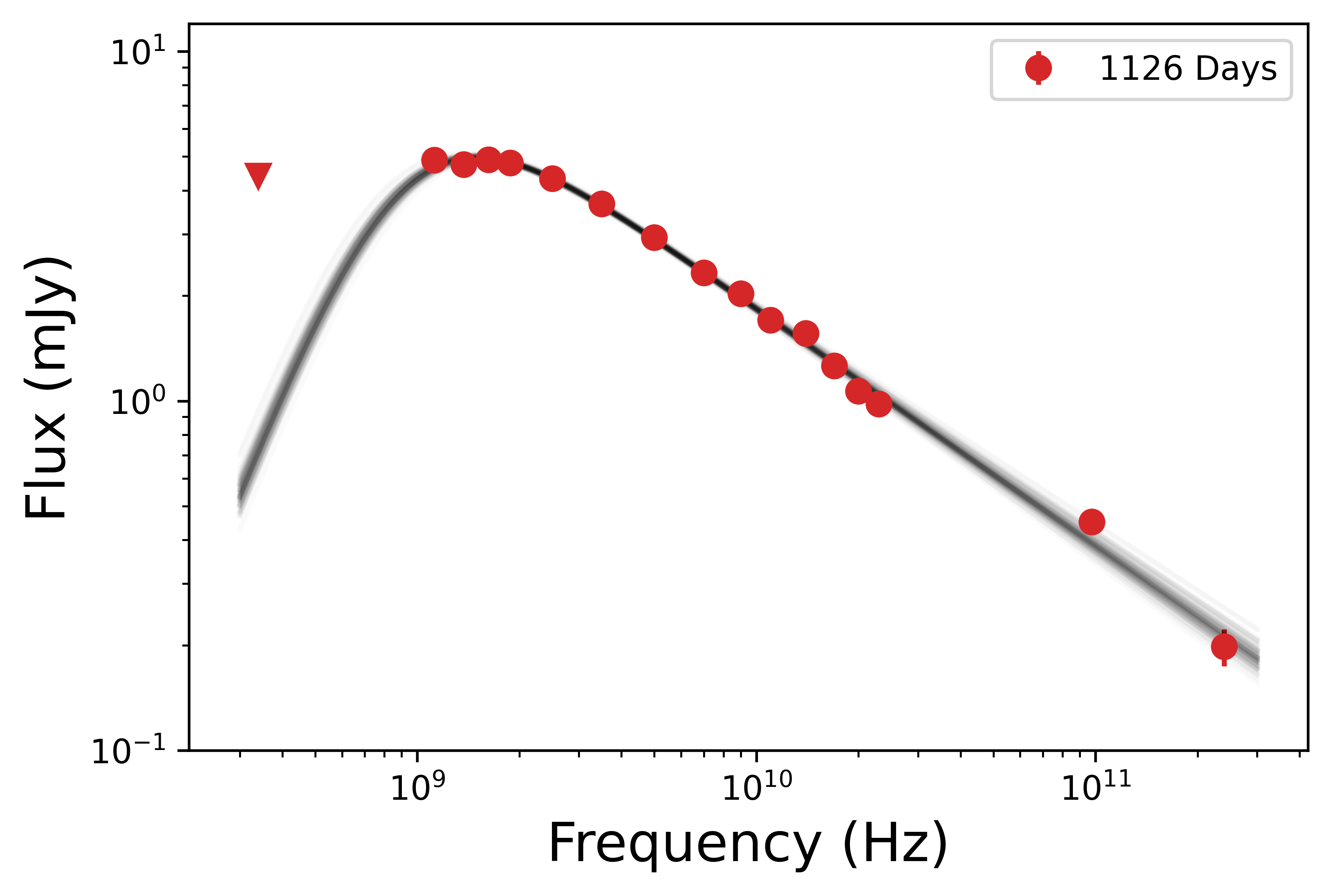}
\includegraphics[width=.45\columnwidth]{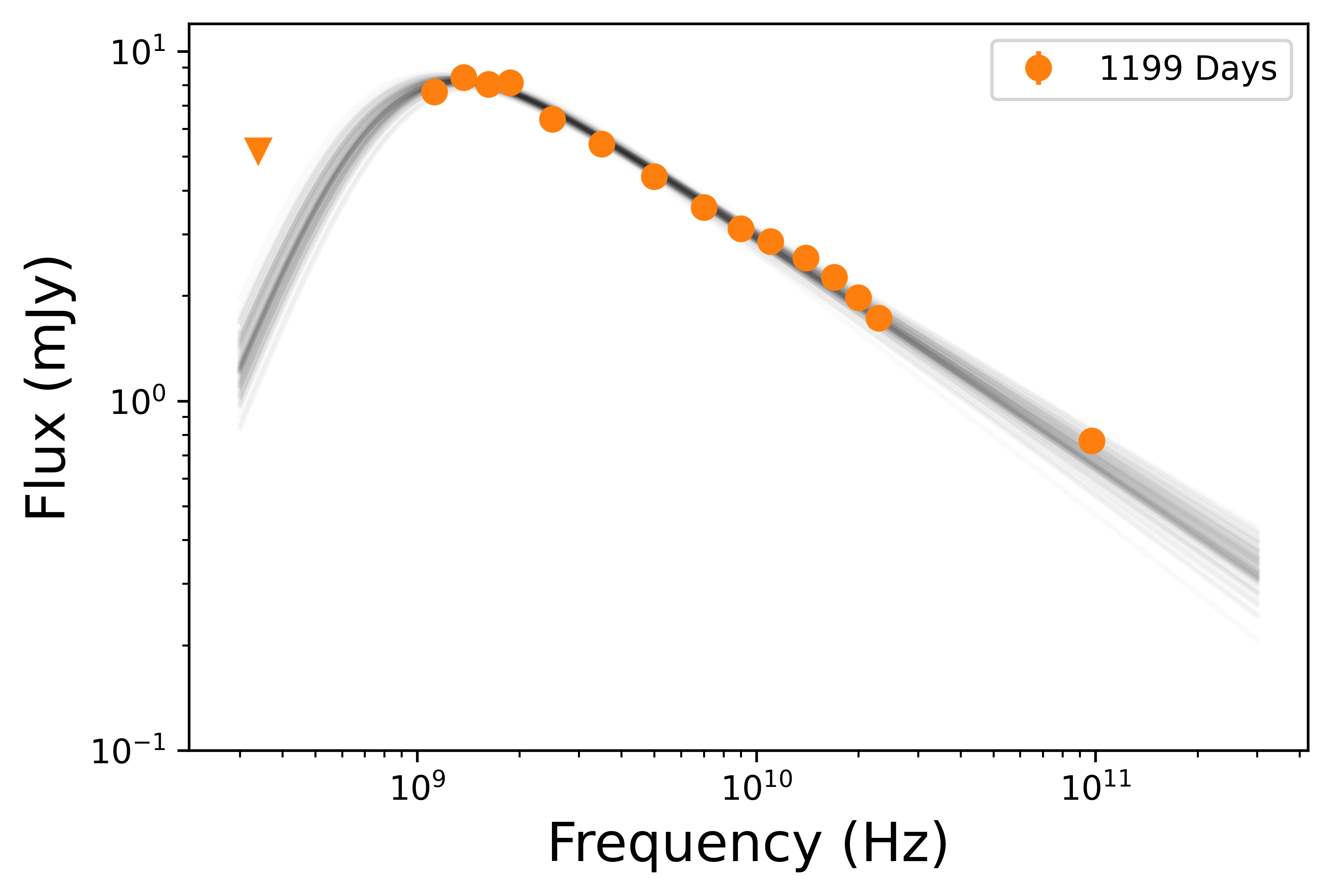}
\includegraphics[width=.45\columnwidth]{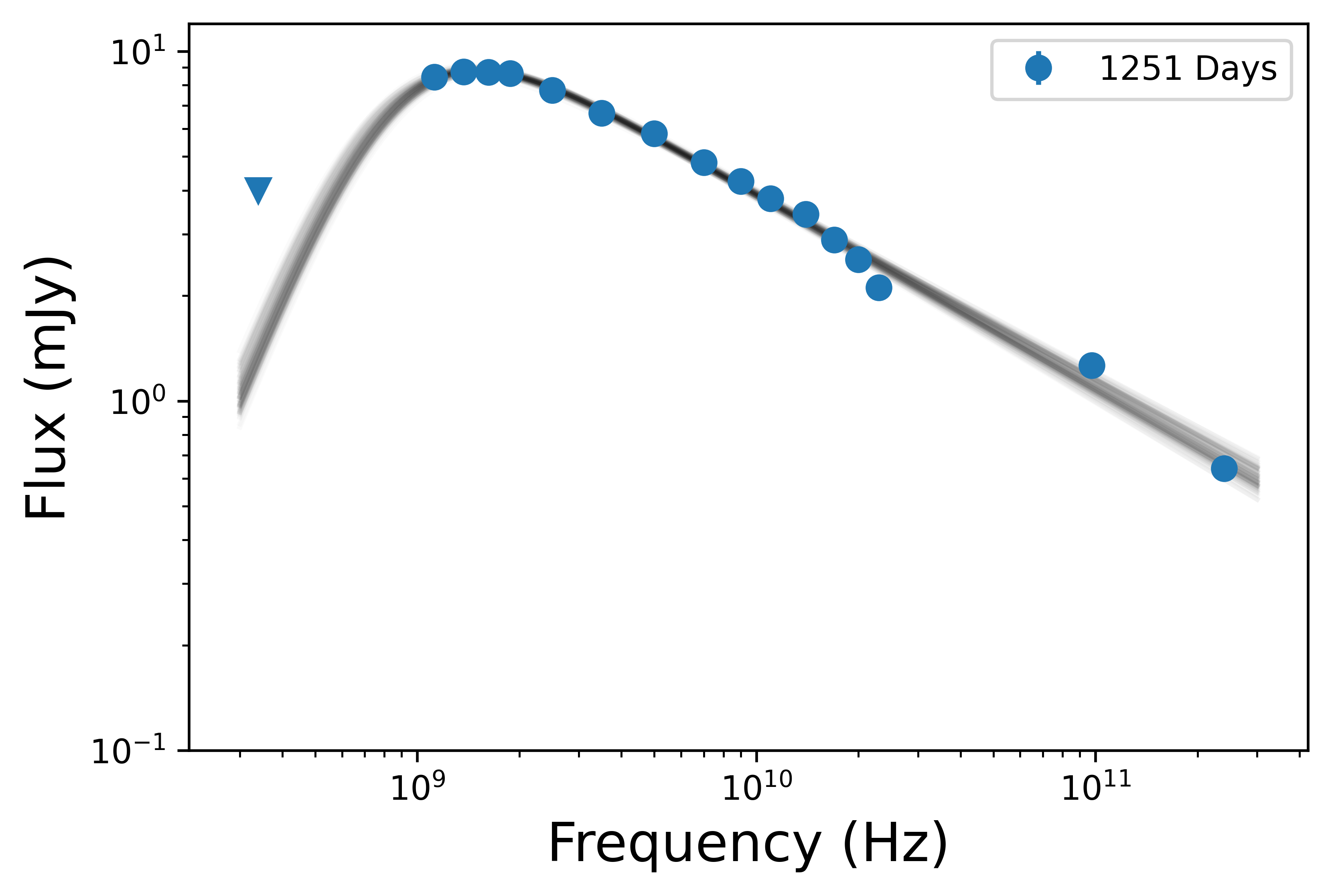}
\includegraphics[width=.45\columnwidth]{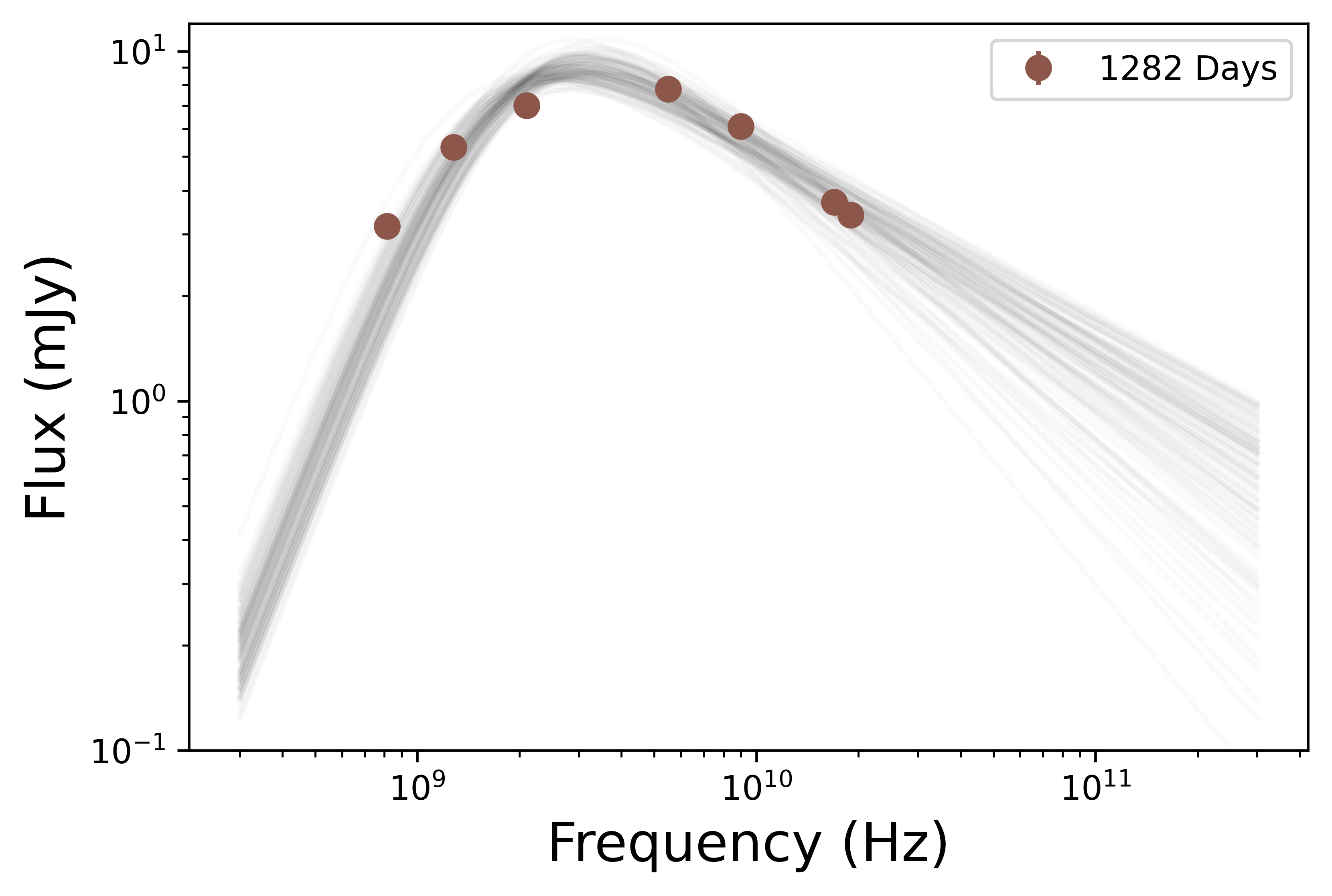}
\end{center}
\label{fig:seds}
\caption{Radio to millimeter spectral energy distributions ranging from 972 days to 1282 days. The grey lines are representative fits from our MCMC modeling (\S\ref{sec:sed}). It is apparent that the first four epochs exhibit a roughly constant peak frequency ($\approx 1.5$ GHz) and a steadily rising peak flux density, while the final epoch exhibits a rapid shift to a higher peak frequency of $\approx 3$ GHz. VLA data at 972 days are combined with VAST data at 1013 days.  VLA data at 1126 days are combined with ALMA data at 1141 days. VLA data at 1199 days are combined with ALMA data at 1201 days.  VLA data at 1251 days are combined with ALMA data at 1253 days. MeerKAT data at 1282 days is combined with ATCA data at 1296 days.}
\end{figure}

\begin{deluxetable*}{lcccc}
\label{tab:sed}
\tablecolumns{5}
\tablecaption{Spectral Energy Distribution Parameters}
\tablehead{
	\colhead{$\delta t$} &
	\colhead{$F_{\nu,p}$} &
	\colhead{log($\nu_p$)} &
	\colhead{$p$} &
	\colhead{$\sigma$} 
	\\
    \colhead{(d)} &
    \colhead{(mJy)} &
    \colhead{(Hz)} &
    \colhead{} &
    \colhead{} 
}
\startdata
972\,$^a$ & $2.38\pm 0.35$ & $9.21\pm 0.05$ & 2.3 & $0.19\pm 0.17$ \\
1126 & $4.98\pm 0.05$ & $9.18\pm 0.01$ & $2.35\pm 0.03$ & $0.06\pm 0.02$\\
1199 & $8.22\pm 0.16$ & $9.12\pm 0.03$ & $2.27\pm 0.06$ & $0.22\pm 0.07$\\
1251 & $8.82\pm 0.10$ & $9.17\pm 0.02$ & $2.09\pm 0.03$ & $0.14\pm 0.03$\\
1282 & $8.83\pm 0.68$ & $9.47\pm 0.04$ & $2.24\pm 0.17$ & $0.74\pm 0.14$\\
\enddata
\tablecomments{$^a$ The values for the first epoch are for a fixed value of $p=2.3$, the average of the first two full SEDs at 1126 and 1199 days.}
\end{deluxetable*}

\section{Modeling and Analysis}
\label{sec:modeling}

\subsection{Modeling of the Radio Spectral Energy Distributions}
\label{sec:sed}

The radio/mm SEDs, shown in Figure~\ref{fig:seds}, exhibit a power law shape with a turnover and peak at $\approx 1.5$ GHz through 1251 days. At 1282 days, however, the peak of the SED shifts upwards to $\approx 3$ GHz.  A rapid shift to a {\it higher} peak frequency is unprecedented in radio observations of TDEs.  The power law shape above the peak is characteristic of synchrotron emission.

We fit the SEDs with the model of \citet{Granot2002}, developed for synchrotron emission from gamma-ray burst (GRB) afterglows, and previously applied to the radio emission from other TDEs (e.g., \citealt{Zauderer2011,Cendes2021}), using specifically the regime\footnote{We note that given our spectral coverage in all but the last observation, and the SED peak at $\approx 1.5$ GHz, we cannot measure the spectral index below the peak to determine if it is $\nu^{5/2}$ (i.e., $\nu_m<\nu_a$) or $\nu^2$ (i.e., $\nu_m>\nu_a$). In the analysis in \S\ref{sec:params} we conservatively assume the former since the latter case would lead to even larger radius and energy, and we find that the resulting parameters indeed support this assumption.} of $\nu_m\ll\nu_a$: 

\begin{equation}
\indent F_\nu = F_\nu (\nu_m) \Bigg[ \Big(\frac{\nu}{\nu_m}\Big)^2 e^{-s_4 (\nu/\nu_m)^{2/3}}+\Big(\frac{\nu}{\nu_m}\Big)^{5/2}\Bigg] 
\times \Bigg[1+\Big(\frac{\nu}{\nu_a}\Big)^{s_5 (\beta_2-\beta_3)}\Bigg]^{-1/s_5},
\label{eq:second-spec}
\end{equation}
where $\beta_2 = 5/2$, $\beta_3 = (1-p)/2$, $s_4 = 3.63p - 1.60$, and $s_5 = 1.25 - 0.18p$ \citep{Granot2002}.  Here, $p$ is the electron energy distribution power law index, $N(\gamma_e)\propto \gamma_e^{-p}$ for $\gamma_e\ge\gamma_m$, $\nu_m$ is the frequency corresponding to $\gamma_m$, $\nu_a$ is the synchrotron self-absorption frequency, and $F_\nu(\nu_m)$ is the flux normalization at $\nu=\nu_m$.

We determine the best fit parameters of the model --- $F_\nu(\nu_m)$, $\nu_a$, and $p$ --- using the Python Markov Chain Monte Carlo (MCMC) module \texttt{emcee} \citep{Foreman-Mackey2013}, assuming a Gaussian likelihood where the data have a Gaussian distribution for the parameters $F_{\nu}(\nu_m)$ and $\nu_a$.  For $p$ we use a uniform prior of $p=2-3.5$. We also include in the model a parameter that accounts for additional systematic uncertainty beyond the statistical uncertainty on the individual data points, $\sigma$, which is this a fractional error added to each data point.  The posterior distributions are sampled using 100 MCMC chains, which were run for 3,000 steps, discarding the first 2,000 steps to ensure the samples have sufficiently converged by examining the sampler distribution. The resulting SED fits are shown in Figure~\ref{fig:seds} and provide a good fit to the data. 

Our first observation at 972 days only includes 5 and 7 GHz, but clearly points to an optically thin spectrum with a peak at $\lesssim 5$ GHz.  We combine this observation with a VAST detection of the source at 1013 days with a flux of 1.3$\pm$0.03 mJy at .89 GHz \citep{Horesh2022}, indicating the peak is between these two frequencies.  We make the reasonable assumption that the lack of evolution in $\nu_p$ between 1126-1251 days indicates no serious change at 972 days as well, and fix $p=2.30$.  We use these values to determine the physical properties of the outflow at 972 days.

From the SED fits we determine the peak frequency and flux density, $\nu_p$ and $F_{\nu,p}$, respectively, which are used as input parameters for the determination of the outflow physical properties.  The best-fit values and associated uncertainties are listed in Table~\ref{tab:sed}. We find that $\nu_p$ remains essentially constant at $\approx 1.3-1.6$ GHz at 972 to 1251 days, while $F_{\nu,p}$ increases steadily by a factor of $3.7$. While the VLITE limits at 350 MHz lie above our SED model fits (Figure~\ref{fig:sed}) we note that they require an SED peak at $\gtrsim 0.5$ GHz since otherwise these limits would be violated.  In addition, the single power law shape above $\nu_p$ indicates that the synchrotron cooling frequency is $\nu_c\gtrsim 240$ GHz at 1126 and 1251 days.  For the SED at 1282 days we find that $F_{\nu,p}$ has remained steady, while $\nu_p$ increased by a factor of 2.  We also note that the spectral index below the peak appears to be shallower than $F_\nu\propto \nu^{5/2}$.

Given the unusual evolution to higher $\nu_p$ in the latest epoch, we have also considered a model with two emission components, peaking at $\nu_{p,1}\approx 1.5$ and $\nu_{p,2}\approx 3$ GHz, in which the lower frequency component dominates the emission at early time, and the higher frequency component rises at later times.  The results of this model are presented in the Appendix (\S\ref{sec:app}), but in the main paper we focus on the simpler single-component model.  

\subsection{Equipartition Analysis}
\label{sec:equi}

Using the inferred values of $\nu_p$, $F_{\nu,p}$ and $p$ from \S\ref{sec:sed}, we can now derive the physical properties of the outflow and ambient medium using an equipartition analysis.  In all epochs, we assume a mean value of  p=2.3 in our calculations.  We first assume the conservative case of a non-relativistic spherical outflow using the following expressions for the radius and kinetic energy \citep[Equations 27 and 28 in ][]{Duran2013}:

\begin{multline} 
R_{\rm eq}\approx (1\times10^{17}\,\, \textrm{cm})\times(21.8\times525^{p-1})^{\frac{1}{13+2p}} \chi_{e}^{\frac{2-p}{13+2p}}F_{p,\rm mJy}^{\frac{6+p}{13+2p}} d_{L,28}^{\frac{2(p+6)}{13+2p}} \nu_{p,10}^{-1}(1+z)^{\frac{-(19+3p)}{13+2p}}
 f_A^{\frac{-(5+p)}{13+2p}} f_V^{\frac{-1}{13+2p}} \\
\times 4^{\frac{1}{13+2p}}  \epsilon^{1/17}\xi^{\frac{1}{13+2p}} \Gamma^{\frac{p+8}{13+2p}} (\Gamma-1)^{\frac{2-p}{13+2p}},
\label{eq:rad}
\end{multline}
\begin{multline} 
E_{\rm eq} \approx (1.3\times10^{48}\,\, \textrm{erg})\times [21.8^{\frac{-2(p+1)}{13+2p}}][525^{p-1}\times \chi_e^{2-p}]^{\frac{11}{13+2p}} F_{p,\rm mJy}^{\frac{14+3p}{13+2p}} d_{L,28}^{\frac{2(3p+14)}{13+2p}} \nu_{p,10}^{-1} (1+z)^{\frac{-(27+5p)}{13+2p}}\\ 
\times f_A^{\frac{-(3(p+1)}{13+2p})} f_V^{\frac{2(p+1)}{13+2p}} 4^{\frac{11}{13+2p}} \xi^{\frac{11}{13+2p}} [(11/17)\epsilon^{-6/17}+(6/17)\epsilon^{11/17}]\Gamma^{\frac{-5p+16}{13+2p}} (\Gamma-1)^{\frac{-11(p-2)}{13+2p}},
\label{eq:energy}
\end{multline}
where $d_L=204$ Mpc is the luminosity distance and $z=0.0457$ is the redshift. The factors $f_A$ and $f_V$ are the area and volume filling factors, respectively, where in the case of a spherical outflow $f_A=1$ and $f_V=\frac{4}{3}\times (1-0.9^3)\approx 0.36$ (i.e., we assume that the emitting region is a shell of thickness $0.1R_{\rm eq}$), while in the case of a jet\footnote{We find here that a collimated outflow is only mildly relativistic, corresponding to the ``narrow jet'' case (\S4.1.1) in \citet{Duran2013}.} $f_A=(\theta_j\Gamma)^2$ and $f_V=0.27(\theta_j\Gamma)^2$.  The factors of $4^{1/13+2p}$ and $4^{11/13+2p}$ in $R_{\rm eq}$ and $E_{\rm eq}$, respectively, arise from corrections to the isotropic number of radiating electrons ($N_{e,\rm iso}$) in the non-relativistic case. We further assume that the fraction of post-shock energy in relativistic electrons is $\epsilon_e=0.1$, which leads to correction factors of $\xi^{1/13+2p}$ and $\xi^{11/13+2p}$ in $R_{\rm eq}$ and $E_{\rm eq}$, respectively, with $\xi=1+\epsilon_e^{-1}\approx 11$.  We parameterize any deviation from equipartition with a correction factor $\epsilon = (11/6)(\epsilon_B/\epsilon_e)$, where $\epsilon_B$ is the fraction of post-shock energy in magnetic fields (here we use $\epsilon_B=0.1$; see \S\ref{sec:cooling}).  Finally, $\chi_e=(p-2)/(p-1)\epsilon_e(m_p/m_e)\approx 42.3$, where $m_p$ and $m_e$ are the proton and electron masses, respectively.

Using $R_{\rm eq}$ we can also determine additional parameters of the outflow and environment \citep{Duran2013}: the magnetic field strength ($B$), the number of radiating electrons ($N_e$), and the Lorentz factor of electrons radiating at $\nu_a$ ($\gamma_a$):
\begin{equation}
B \approx (1.3\times10^{-2}\, \textrm{G})\times  F_{p,\rm mJy}^{-2} d_{L,28}^{-4} (1+z)^{7}  \nu_{p,10}^5 \frac{f_A^{2} R_{\rm eq,17}^{4}}{\Gamma^3}
\label{eq:mag}
\end{equation}
\begin{equation}
N_{e} \approx (4\times10^{54})\times F_{p,\rm mJy}^{3} d_{L,28}^{6} (1+z)^{-8} f_A^{-2} R_{\rm eq,17}^{-4} \nu_{p,10}^{-5} \times (\gamma_{a}/\gamma_{m})^{(p-1)},
\label{eq:dens}
\end{equation}
\begin{equation}
\gamma_a \approx 525\times F_{p,\rm mJy} d_{L,28}^2 \nu_{p,10}^{-2} (1+z)^{-3} \frac{\Gamma}{f_A R_{\rm eq,17}^{2}}.
\label{eq:gamma}
\end{equation}
We note an additional factor of 4 and a correction factor of $(\gamma_{a}/\gamma_{m})^{p-1}$ are added to $N_e$ for the non-relativistic regime (Barniol Duran, private communication); here we use $\gamma_m={\rm max}[\chi_e(\Gamma-1),2]$. We determine the ambient density assuming a strong shock and an ideal monoatomic gas as $n_{\rm ext} = N_e/4V$, where the factor of 4 is due to the shock jump conditions and $V$ is the volume of the emitting region, $V=f_V\pi R_{\rm eq}^3/\Gamma^4$, with $f_V$ as defined above.

\begin{deluxetable*}{l|llllllllll}
\label{tab:equi}
\tablecolumns{11}
\tablecaption{Equipartition Model Parameters}
\tablehead{
	\colhead{Geometry} &
	\colhead{$\delta t$} &
	\colhead{log($R_{\rm eq}$)} &
	\colhead{log($E_{\rm eq}$)} &
	\colhead{log($B$)} &
	\colhead{log($N_e$)} &
	\colhead{log($n_{\rm ext}$)} &
	\colhead{$\Gamma$} &
	\colhead{$\beta$} &
	\colhead{$\gamma_a$} &
	\colhead{$\nu_c$} 
	\\
	\colhead{} &
    \colhead{(d)} &
    \colhead{(cm)} &
    \colhead{(erg)} &
    \colhead{(G)} &
    \colhead{} &
    \colhead{(cm$^{-3}$)} &
    \colhead{} &
    \colhead{} &
    \colhead{} &
    \colhead{(GHz)} 
}
\startdata
 spherical  & 972 &$17.22\substack{+0.09 \\ -0.06}$&$49.03\substack{+0.13 \\ -0.09}$&$-0.79\substack{+0.05 \\ -0.07}$&$52.54\substack{+0.13 \\ -0.09}$&$0.27\substack{+0.10 \\ -0.14}$&$1.03\substack{+0.01 \\ -0.01}$&$0.23\substack{+0.03 \\ -0.03}$ & $70\substack{+1 \\ -1} $& $1061\substack{+1039 \\ -1212}$ \\
($f_A=1,$ & 1126 &$17.41\substack{+0.01 \\ -0.01}$&$49.47\substack{+0.01 \\ -0.01}$&$-0.85\substack{+0.01 \\ -0.01}$&$52.96\substack{+0.01 \\ -0.01}$&$0.12\substack{+0.03 \\ -0.03}$&$1.02\substack{+0.01 \\ -0.01}$&$0.22\substack{+0.01 \\ -0.01}$& $63 \substack{+1 \\ -1}$ &$579\substack{+48 \\ -62}$ \\
$f_V=0.36,$ & 1199 &$17.57\substack{+0.03 \\ -0.03}$&$49.76\substack{+0.03 \\ -0.03}$&$-0.96\substack{+0.03 \\ -0.03}$&$53.28\substack{+0.03 \\ -0.03}$&$-0.04\substack{+0.06 \\ -0.06}$&$1.03\substack{+0.01 \\ -0.01}$&$0.25\substack{+0.01 \\ -0.01}$& $65 \substack{+1 \\ -1}$ &$838\substack{+158 \\ -209}$\\
$\epsilon_B=0.1)$ & 1251 &$17.54\substack{+0.02 \\ -0.02}$&$49.75\substack{+0.02 \\ -0.02}$&$-0.91\substack{+0.02 \\ -0.02}$&$53.30\substack{+0.02 \\ -0.02}$&$-0.02\substack{+0.04 \\ -0.03}$&$1.02\substack{+0.01 \\ -0.01}$&$0.22\substack{+0.01 \\ -0.01}$& $65 \substack{+1 \\ -1}$ &$492\substack{+62 \\ -77}$\\
& 1282 &$17.24\substack{+0.04 \\ -0.04}$&$49.57\substack{+0.04 \\ -0.05}$&$-0.56\substack{+0.05 \\ -0.05}$&$52.92\substack{+0.04 \\ -0.05}$&$0.54\substack{+0.09 \\ -0.10}$&$1.01\substack{+0.01 \\ -0.01}$&$0.12\substack{+0.01 \\ -0.01}$& $62 \substack{+1 \\ -1}$ &$377\substack{+165 \\ -122}$\\
\hline
$10^{\circ}$ jet  & 972 &$17.89\substack{+0.09 \\ -0.06}$&$49.09\substack{+0.13 \\ -0.09}$&$-1.03\substack{+0.05 \\ -0.07}$&$52.65\substack{+0.13 \\ -0.09}$&$0.02\substack{+0.11 \\ -0.14}$&$1.20\substack{+0.04 \\ -0.04}$&$0.55\substack{+0.05 \\ -0.04}$& $96 \substack{+1 \\ -1}$ &$5465\substack{+1034 \\ -1196}$ \\
($f_A=\theta^2\Gamma^2,$ & 1126 &$18.08\substack{+0.01 \\ -0.01}$&$49.52\substack{+0.01 \\ -0.01}$&$-1.09\substack{+0.01 \\ -0.01}$&$53.09\substack{+0.01 \\ -0.01}$&$-0.13\substack{+0.03 \\ -0.03}$&$1.21\substack{+0.01 \\ -0.01}$&$0.56\substack{+0.01 \\ -0.01}$& $76 \substack{+1 \\ -1}$ &$2965\substack{+247 \\ -318}$ \\
$f_V=0.27f_A)$ & 1199 &$18.24\substack{+0.03 \\ -0.03}$&$49.81\substack{+0.03 \\ -0.03}$&$-1.20\substack{+0.03 \\ -0.03}$&$53.36\substack{+0.03 \\ -0.03}$&$-0.29\substack{+0.06 \\ -0.06}$&$1.26\substack{+0.02 \\ -0.02}$&$0.61\substack{+0.02 \\ -0.02}$& $78 \substack{+1 \\ -1}$ &$4341\substack{+819 \\ -1082}$\\
$\epsilon_B=0.1)$ & 1251 &$18.21\substack{+0.02 \\ -0.02}$&$49.80\substack{+0.02 \\ -0.02}$&$-1.15\substack{+0.02 \\ -0.02}$&$53.37\substack{+0.02 \\ -0.02}$&$-0.24\substack{+0.04 \\ -0.04}$&$1.21\substack{+0.01 \\ -0.01}$&$0.57\substack{+0.01 \\ -0.01}$& $78 \substack{+1 \\ -1}$ &$2523\substack{+322 \\ -388}$\\
& 1282 &$17.93\substack{+0.04 \\ -0.04}$&$49.60\substack{+0.04 \\ -0.05}$&$-0.79\substack{+0.05 \\ -0.05}$&$53.18\substack{+0.04 \\ -0.05}$&$0.31\substack{+0.09 \\ -0.10}$&$1.09\substack{+0.01 \\ -0.01}$&$0.39\substack{+0.02 \\ -0.01}$& $79 \substack{+1 \\ -1}$ &$188\substack{+56 \\ -72}$\\
 \enddata
 \tablecomments{Values in this table are calculated using an outflow launch time of $t_0=750$ days.  For $\Gamma$ and $\beta$ we have accounted for the uncertainty in the launch date ($t_{\rm 0,sphere}=750\substack{+80 \\ -127}$ days and $t_{\rm 0,jet}=750\substack{+73 \\ -113}$).}
\end{deluxetable*}

\subsection{Cooling Frequency and $\epsilon_{B}$}
\label{sec:cooling}

\begin{figure}[t!]
\begin{center}
        \includegraphics[width=0.65\columnwidth]{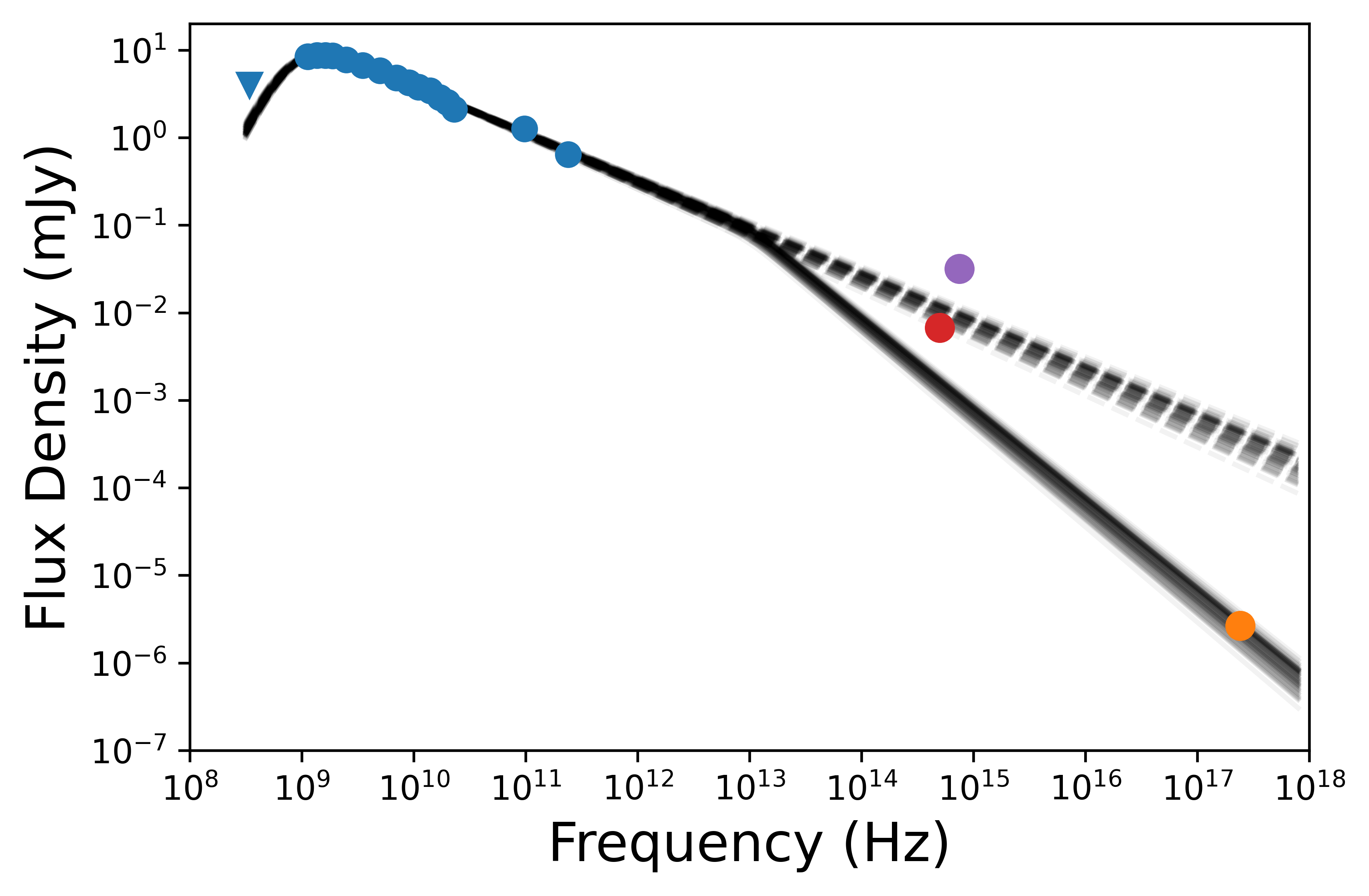}
        \end{center}
    \caption{VLA+ALMA data at at $1251-1253$ days (blue points), \textit{Chandra} data at 1253 days (yellow point), UVOT data at 1182 days (purple circle), and ZTF data at 1030 days (red circle).  Also shown are the models from Figure~\ref{fig:sed} (dashed; no cooling break), and models that include a cooling break to match the X-ray flux density (black; Equation~\ref{eq:cooling}).  We find that $\nu_c\approx 10^{13}$ Hz.}
    \label{fig:cooling}
\end{figure}

Using the inferred physical parameters we can also predict the location of the synchrotron cooling frequency, given by \citep{Sari1998}:
\begin{equation} 
\nu_c\approx 2.25\times 10^{14} B^{-3} \Gamma^{-1} t_d^{-2}\,\,{\rm Hz}.
\label{eq:cooling}
\end{equation}
Since $\nu_c$ has a strong dependence on the magnetic field strength, measuring its location directly can determine $\epsilon_B$ and whether the outflow deviates from equipartition.
 
In Figure~\ref{fig:cooling} we show our VLA+ALMA+\textit{Chandra} SED at $1251-1253$ days, along with our model SED from \S\ref{sec:sed}, which does not include a cooling break (dashed lines). This model clearly over-predicts the \textit{Chandra} measurements.  The steepening required by the \textit{Chandra} data is indicative of a cooling break, which we model with an additional multiplicative term to Equation~\ref{eq:second-spec} of
$[1+({\nu/\nu_c})^{s_3(\beta_3-\beta_4)}]^{-1/s_3}$,
where $\beta_4 = -p/2$ and we use $s_3=10$ \citep{Granot2002,Cendes2021b}.  Fitting this model to the data we find $\nu_c\approx 10^{13}$ Hz (solid lines in Figure~\ref{fig:cooling}). However, since the X-ray flux measured with \textit{Chandra} is only a factor of 2 times fainter than the steady early-time X-ray flux, and hence can be due to a source other than the radio-emitting outflow, we consider it to be an upper limit on the contribution of the radio-emitting outflow.  As a result, our estimate of $\nu_c$ is actually an upper limit.

With the value of $\nu_c$ determined, we adjust the value of $\epsilon_B$ and solve Equation~\ref{eq:cooling} after repeating the equipartition analysis (Equations~\ref{eq:rad} to \ref{eq:dens}) to account for the deviation from equipartition in those parameters.  With this approach, we find that $\epsilon_{B}\approx 0.01$. Given that the deviation is not significant, and that $\nu_c$ is an upper limit (and hence $\epsilon_B\approx 0.01$ is a lower limit) we conservatively assume equipartition ($\epsilon_B=0.1$) in our subsequent analysis and in Table \ref{tab:equi}.  We emphasize that the change would be relatively minor if we adjusted $\epsilon_{B}$- at 1251 days, for the jetted model we find this would correspond with a radius decrease from ${\rm log}(R_{\rm eq})\approx 18.21$ to $\approx 18.16$, and the energy would increase from ${\rm log}(E_{\rm eq})\approx 49.80$ to $\approx 49.98$.

\section{Physical Properties of the Outflow}
\label{sec:params}

\begin{figure}
\begin{center}
    \includegraphics[width=.75\columnwidth]{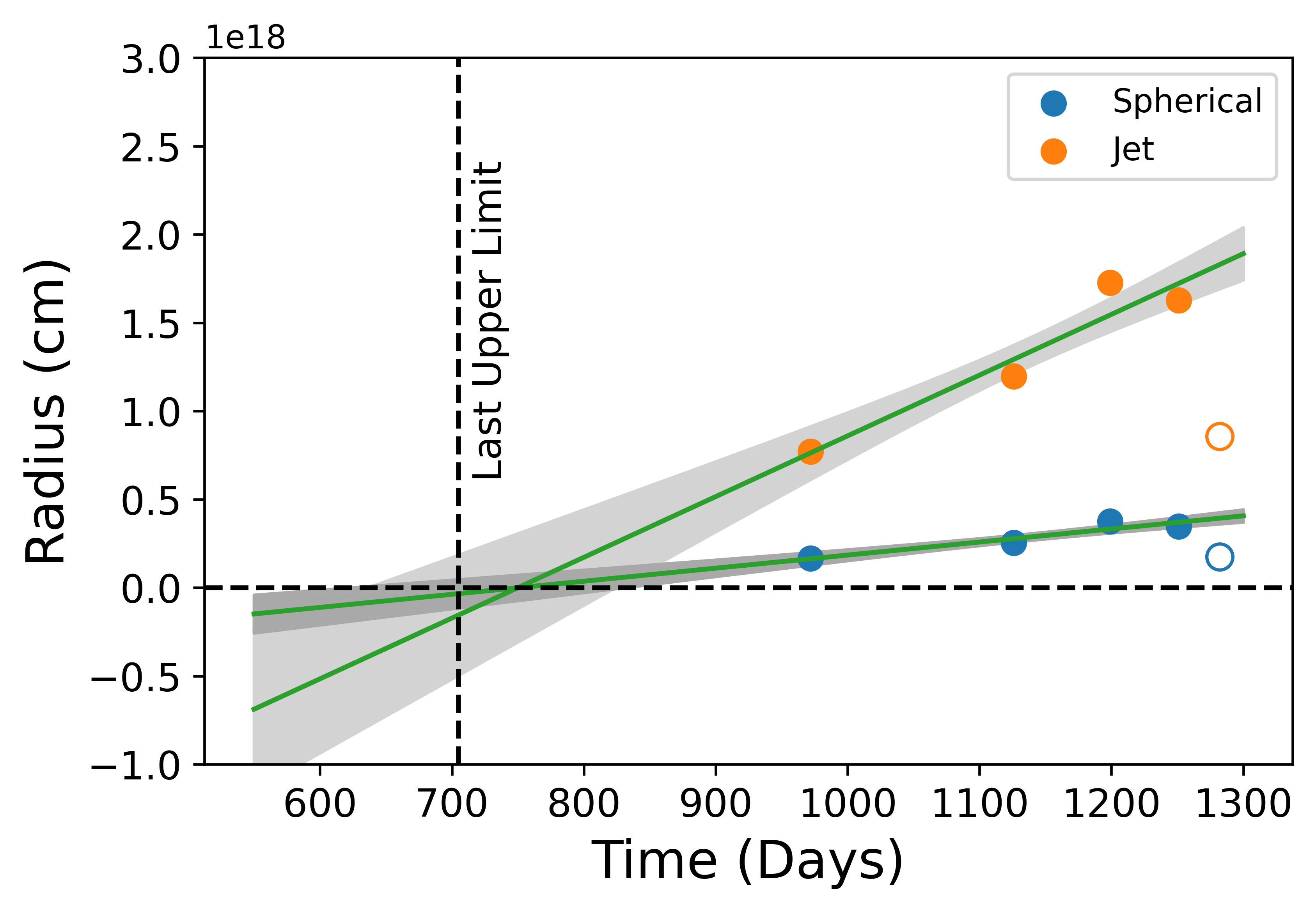}
    \end{center}
    \label{fig:disruption}
    \caption{Radius evolution of the outflow for the spherical (blue) and jet (orange) geometries (Table~\ref{tab:equi}), assuming equipartition.  We fit a linear trend (i.e., free expansion) to these data to determine the launch time of the outflow (green lines) and its uncertainty (grey shaded regions mark the $1\sigma$ range), excluding the observation at 1282 days (open circle) for reasons outlined in Section \ref{sec:modeling}. We find that $t_{\rm 0,sphere}=750\substack{+80 \\ -127}$ days and $t_{\rm 0,jet}=750\substack{+73 \\ -113}$ days relative to the time of optical discovery.  We also mark the time of the final non-detection at $705$ days for reference (vertical dashed line).}
\end{figure}

\subsection{Spherical Outflow}
\label{sec:spherical}

We begin by investigating the properties of the outflow in the conservative case of spherical geometry.  We summarize the inferred physical parameters for all epochs in Table~\ref{tab:equi}. We find that the radius increases from ${\rm log}(R_{\rm eq})\approx 17.22$ to $\approx 17.57$ between 970 and 1251 d, corresponding to a large velocity of $\beta\approx 0.28$ over this time span.  However, if we use the time of optical discovery as the outflow launch date, then the inferred velocity in the first epoch (972 d) is $\beta\approx 0.066$ and in the fourth epoch (1251 d) it is $\beta\approx 0.11$.  This means that the assumption of a launch date that coincides with the optical discovery is incorrect.  Instead, we find that the increase in radius during the first four epochs is roughly linear, and fitting such an evolution with the launch time (i.e., time at which $R=0$) as a free parameter, we instead find $t_0\approx 750$ d; see Figure \ref{fig:disruption}.  Thus, the physical evolution of the radius during this period confirms our initial argument based on the rapid rise of the radio emission (\S\ref{sec:lumin}) that the outflow was launched with a substantial delay of about 2 years relative to the optical emission. The inferred outflow velocity using the four epochs is $\beta\approx 0.24$, which is larger than the typical velocities of $\beta\approx 0.1$ inferred for previous non-relativistic radio-emitting TDEs \citep{Cendes2021b,Goodwin2022,Alexander2016,Alexander2017,Anderson2019}.  

The outflow kinetic energy increases by a factor of about 5 from $E_K\approx 1.1\times 10^{49}$ to $5.8\times 10^{49}$ erg between 970 and 1250 days.  The rapid increase in energy indicates that we are observing the deceleration of the outflow.  The kinetic energy is larger than that of previous radio-emitting TDEs by a factor of $\approx 4$ \citep{Goodwin2022,Anderson2019}, although we note that this energy is a lower limit and would be higher if a deviation from equipartition is assumed. 

Finally, we find that the inferred ambient density declines from $n_{\rm ext}\approx 1.9$ to $\approx 1.0$ cm$^{-3}$ over the distance scale of $1.7\times 10^{17}$ to $3.7\times 10^{17}$ cm, or $\rho(R)\propto R^{-1}$.  This is similar to the rate seen in M87* and Sw\,J1644+57 and Sgr A*, and less steep than the density profiles inferred around previous thermal TDEs (see Figure \ref{fig:density} and citations therein).  Combined with the mild decline in density with radius, this indicates that the late turn-on of the radio emission is inconsistent with a density jump.  

To conclude, even in the conservative spherical scenario we find that radio emission requires a delayed, mildly-relativistic outflow with a higher velocity and energy than in previous radio-emitting TDEs.  

\subsection{Collimated Outflow}

In light of the large outflow velocity and kinetic energy in the spherical case, we also consider the results for a collimated outflow. In particular, we choose an outflow opening angle of $10^\circ$, typical of GRB jets and the jet in Sw\,J1644+57 \citep{p1,p2}. For a collimated outflow the resulting radius (and hence velocity) are larger than in the spherical case, so we need to also solve for the Lorentz factor, $\Gamma$, which impacts the values of $R_{\rm eq}$ (as well as the other physical parameters). We begin by using the launch time inferred in the spherical case, and then iteratively re-calculate $\Gamma$, $R_{\rm eq}$, and the launch date. In the process we also include the relevant modifications due to $\Gamma$ in $f_A$ and $f_V$, which also impact the value of $R_{\rm eq}$ (Equation~\ref{eq:rad}). We note that these corrections are relatively small but we account for them for completeness.

With this approach, we find that the launch date for a $10^\circ$ jet is $t_0\approx 750$ d post optical discovery.  The resulting radii are a factor of 4.7 times larger than in the spherical case, leading to a mean velocity to 1251 days of $\beta\approx 0.57$, or $\Gamma\approx 1.23$.  The kinetic energy is $E_K\approx 6.3\times10^{49}$ erg at 1251 days. Finally, we find the density declines from $n_{\rm ext}\approx 1.0$ to $\approx 0.5$ cm$^{-3}$ over the distance scale of $7.8\times 10^{17}$ to $1.7\times 10^{18}$ cm.  This is less dense than the spherical case, but follows a similar profile of $\rho(R)\propto R^{-1}$ (\S\ref{sec:spherical}).


\subsection{Off-Axis Jet and Other Scenarios}

We can also consider the possibility that the late-time emission from AT2018hyz is caused by a relativistic jet with an initial off-axis viewing orientation. The radio emission from an off-axis relativistic jet will be suppressed at early times by relativistic beaming, but will eventually rise rapidly (as steep as $t^3$) when the jet decelerates and spreads.  The time and radius at which the radio emission will peak are given by (e.g., \citealt{Nakar2011}):
\begin{equation}
t_{\rm dec} \approx 30\, {\rm d}\,\, E_{\rm eq,49}^{1/3} n_{\rm ext}^{-1/3} \beta_0^{-5/3}\\
\label{eq:tdec}
\end{equation}
\begin{equation}
R_{\rm dec} \approx 10^{17}\, {\rm cm}\,\, E_{\rm eq,49}^{1/3} n_{\rm ext}^{-1/3}\\ \beta_0^{-2/3},
\label{eq:rdec}
\end{equation}
where $\beta_0$ is the initial velocity, which for an off-axis relativistic jet is $\beta_0=1$. Using the equipartition parameters in the spherical case, we find $t_{\rm dec}\approx 25$ d and $R_{\rm dec}\approx 8.3\times 10^{16}$ cm.  The value of $t_{\rm dec}$ is substantially smaller than the observed delay of $\sim 10^3$ d (as is $R_{\rm dec}$). This agrees with our general argument in \S\ref{sec:lumin} that the lower radio luminosity of AT2018hyz and its much later appearance compared to the radio emission of Sw\,J1644+57 (which became non-relativistic at $\approx 700$ d) argues against an off-axis jet launched at the time of disruption.  Moreover, the radio emission from an off-axis jet is expected to rise no steeper than $t^3$ \citep{Nakar2011}, whereas in \tde\ the emission rises as $t^5$ if the outflow was launched at the time of disruption.

We can consider the possibility of an off-axis jet expanding initially into a low-density cavity, followed by a denser region (thus delaying $t_{\rm dec}$ to a longer timescale as observed).  However, we can rule out this model because the observed rise in radio emission spans $\approx 300$ d, while Equation~\ref{eq:tdec} indicates deceleration over a timescale of $t_{\rm dec}\approx 30$ d even if the time at which deceleration starts is itself delayed.  Similarly, if there was initially a higher density environment which we did not capture in our observations, this would only correspond with a faster $t_{\rm dec}$.

We also consider the hypothesis that the rise in radio emission is caused by unbound material from the initial disruption colliding with a surrounding dense circumstellar material (CSM).  Theoretical modeling of such unbound debris indicates the fastest speeds reached are $v\approx 0.03c$ \citep{Yalinewich2019,Guillochon2016}, which is significantly smaller than what we infer in both the spherical and jetted cases for \tde. We thus conclude this scenario is unlikely.

Finally, we considered a model in which the change in the SED properties during the latest epoch could be due to a combination of two outflows, with one dominating at a lower frequency and fading and the second dominating at higher frequencies and rising (see Appendix). Such a model may be expected if internal shocks within the outflow are leading to dissipation at more than one radius.  However, we find that such a model requires the initial emission component to decline very rapidly (effectively turn off), while the later emission component has to rise by more than an order of magnitude in only 30 days between about 1250 and 1280 days. Such rapid evolution does not seem feasible, even in the context of \tde.

\section{Discussion and Comparison to Previous Radio-Emitting TDEs}
\label{sec:disc}

\begin{figure}
\begin{center}
    \includegraphics[width=.9\columnwidth]{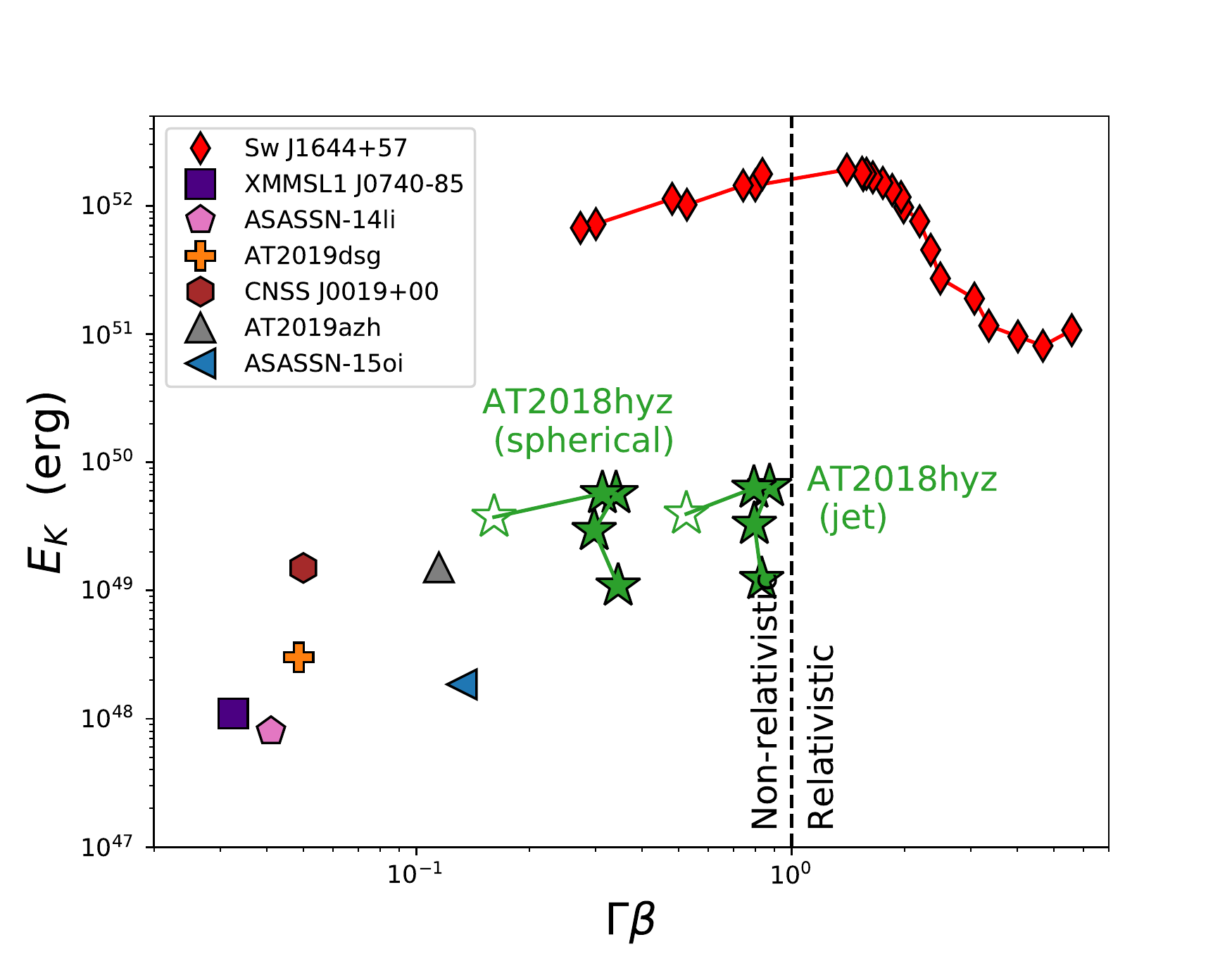}
    \end{center}
    \label{fig:Ev}
    \caption{The energy/velocity for AT2018 spherical case and jetted case; note in both cases the energy increases over time.  We include non-relativistic TDEs assuming a spherical outflow \citep{Cendes2021b,Stein2020,Alexander2016, Alexander2017, Anderson2019,Goodwin2022}, and Sw\,J1644+57 which launched a relativistic jet \citep{Zauderer2011,Cendes2021}.  In the case of AT2019dsg, we display the highest energy inferred in the system.  In the case of AT2019azh, which showed significant fluctuations in the final energy due to changes in $p$ in some later observations \citep{Goodwin2022}, we show the energy and velocity at the peak flux in the luminosity curve adjusted to $\epsilon_{B} = 0.1$.  For ASASSN-15oi, we use the observation with the highest peak frequency and peak flux (182 days) $\epsilon_{B} = 0.1$ and $p=2.39$, which best fit the SED at the highest peak flux observation at 182 days post-disruption, and infer the velocity by subtracting 90 days \citep[the last date of non-detection; see][]{Horesh2021}. We find AT2018hyz is more energetic than the non-relativistic outflow TDEs, and has a higher velocity.}
\end{figure}

\subsection{Outflow Kinetic Energy and Velocity}

In Figure~\ref{fig:Ev} we plot the kinetic energy and velocity ($\Gamma \beta$) of the delayed outflow in \tde\ in comparison to previous TDEs for which a similar analysis has been carried out, using the highest energy inferred in those sources \citep{Cendes2021b,Stein2020,Alexander2016, Alexander2017, Anderson2019,Goodwin2022,Zauderer2011,Cendes2021}. We find that at its peak, in the spherical case the energy is $\approx 4$ times larger and the velocity is $\approx 2$ times faster than in previous non-relativistic TDEs.  If we compare to ASASSN-15oi specifically, using the observation with the highest peak frequency and peak flux (i.e., 182 days post optical discovery), with $\epsilon_B = 0.1$ and $p=2.4$ \citep[which best fits the SED; see][]{Horesh2021}, we obtain an energy for ASASSN-15oi that is $\approx 23$ times lower than in \tde{}.  To infer the velocity in ASASSN-15oi, we subtract 90 days from the date of the observation \citep[the last date of non-detection; see][]{Horesh2021} to find $\beta\approx 0.13$, which is lower than $\beta\approx 0.25$ for \tde.

If the outflow in \tde{} is collimated, then its velocity is significantly higher than in previous non-relativistic TDEs, placing it in an intermediate regime with the powerful jetted TDEs such as Sw\,J1644+57.

\subsection{Circumnuclear Density}

\begin{figure}
\begin{center}
    \includegraphics[width=.9\columnwidth]{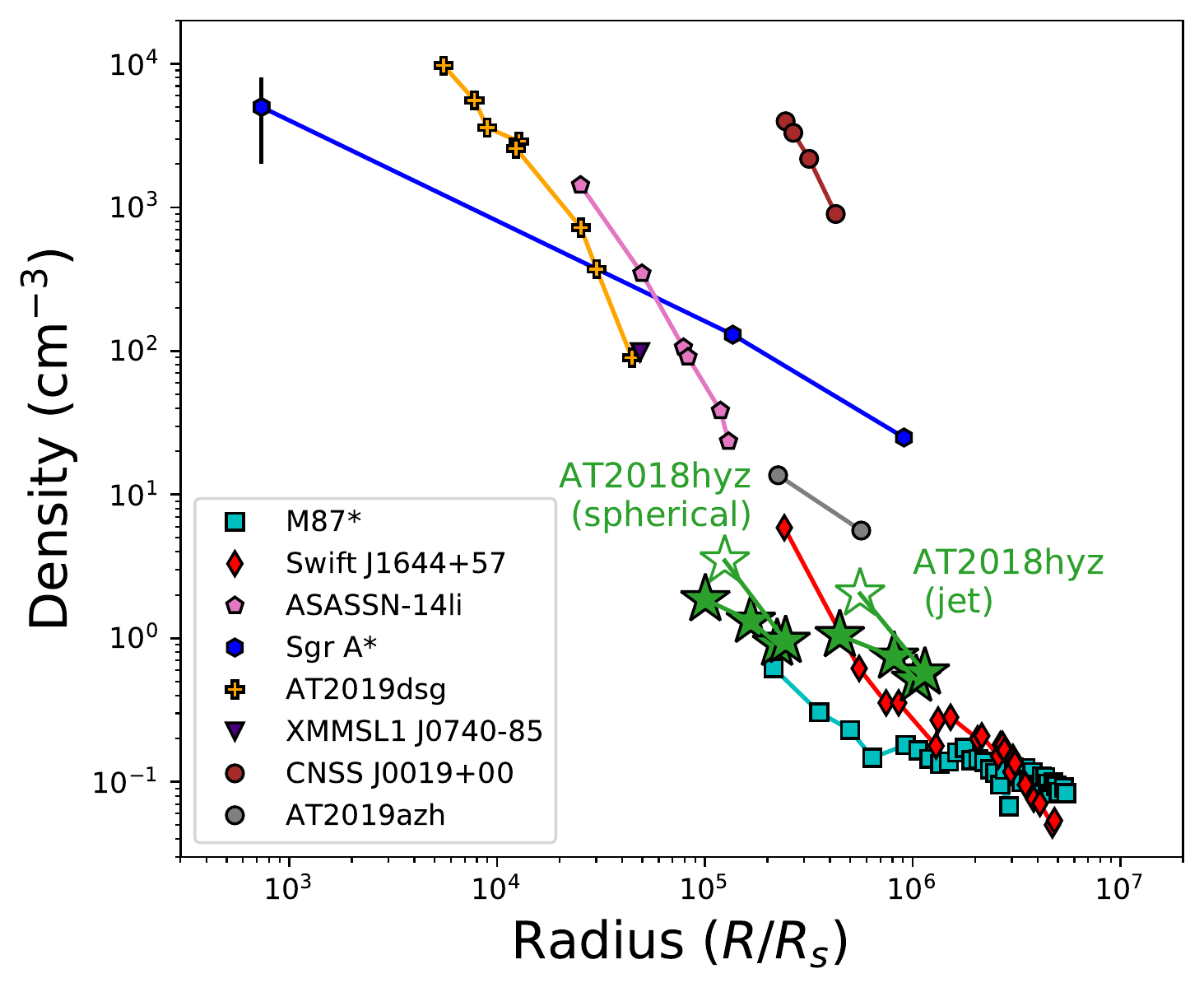}
    \end{center}
    \label{fig:density}
    \caption{The circumnuclear density profile derived from various TDEs including \tde, normalized to the Schwarzschild radius of the SMBH at each host galaxy's center. \tde's host galaxy is lower density, similar to that seen in the jetted TDE SwJ1644+57 \citep{p1,p2,Eftekhari2018} and M87 \citep{Russell2015}. We also include non-relativistic TDEs (e.g.~ASASSN-14li,  \citealt{Alexander2016}, AT2019dsg \citealt{Cendes2021b}, AT2019azh \citep{Goodwin2022}, ASASSN-15oi \citep{Horesh2021}), and for the Milky Way \citep{Baganoff2003,Gillessen2019}.  Other than for AT2019dsg and Sw\,J1644+57, where the cooling break is detected, we assume equipartition in all radio TDEs, so their densities are lower limits.}
\end{figure}

In Figure~\ref{fig:density} we plot the ambient density as a function of radius (scaled by the Schwarzschild radius) for \tde\ and previous radio-emitting TDEs.  Here we use $M_{\rm BH}\approx 5.2\times 10^{6} M_{\odot}$ for \tde, as inferred by \citet{Gomez2020}.  We find that the density decreases with radius, and is consistent with the densities and circumnuclear density profiles of previous TDEs, including Sw\,J1644+57  Crucially, we do not infer an unusually high density, which might be expected if the radio emission was delayed due to rapid shift from low to high density.

\subsection{Comparison to Other TDEs with Late Radio Emission}

Two previous TDEs with delayed radio emission have been published recently. The radio emission from iPTF16fnl is detected on a much earlier timescale than in \tde\ and rises more gradually to a peak luminosity of only $\approx 10^{37}$ erg s$^{-1}$ \citep{Horesh2021b}.  This is almost a factor of 100 less luminous than \tde. We do not consider the radio emission from iPTF16fnl to be similar in nature to \tde.

On the other hand, ASASSN-15oi exhibits two episodes of rapid brightening, at $\approx 200$ and $\approx 1400$ d \citep{Horesh2021}.  While the second brightening is not well characterized temporally or spectrally, it has a comparable rise rate and luminosity to \tde.  We speculate that it may be due to a delayed outflow with similar properties to that of \tde, including a delay of several hundred days, which would make it distinct from the first peak in the ASASSN-15oi light curve.

\subsection{Origin of the Delayed Outflow}

There are at least two broad possibilities for the origin of the delayed mildly relativistic outflow, both of which connect to its assumed origin in a fast disk wind or jet (hereafter, collectively referred to as jet) from the  innermost regions of the black hole accretion flow.  

One possibility is that the jet was weak or inactive at early times after the disruption and then suddenly became activated at $\approx 750$ d.  Such sudden activation could result from a state change in the SMBH accretion disk, such as a thin disk that transitioned to a hot accretion flow.  This is predicted to occur -- in analogy with models developed to explain state changes in X-ray binaries -- when the mass accretion rate falls below a few percent of the Eddington rate (e.g., \citealt{Tchekhovskoy2014}).  The optical light curve of \tde\ extends to $\approx 800$ d \citep{Gomez2020,Hammerstein2022,Short2020}, overlapping the time at which we estimate the outflow was launched. When examining the data from ZTF at later times, we find there is an excess (\S\ref{sec:obs}).  Based on analytical modeling of the optical data using the MOSFiT code \citep{Gomez2020} we estimate that the mass accretion rate at the time the radio outflow was launched is $\approx 0.05 \dot{M}_{\rm Edd}$, consistent with the possibility of a state change.

An alternative explanation is that powering a relativistic jet via the Blandford-Znajek process requires a strong magnetic flux threading the black hole horizon.  The original magnetic field of the disrupted star in a TDE is not expected to contain a strong enough magnetic field to power a relativistic jet \citep{Giannios11}, which requires an alternative origin. The first possibility is that the magnetic flux could be generated through a dynamo by the accretion disk itself; \citet{Liska2020} found that it may take only $\sim 10$ days to generate poloidal flux from the toroidal field through a dynamo effect once the disk is sufficiently thick, thereby connecting jet production to the disk getting thinner as the accretion rate drops.
Alternatively, \citet{Tchekhovskoy2014} and \citet{Kelley2014} suggest that the required magnetic flux may originate from a pre-existing AGN disk, which is `lassoed' in by the infalling fall-back debris; because the matter falling back at later and later times in a TDE reaches larger and larger apocenter radii, depending on the radial profile of the magnetic flux in the pre-existing AGN disk, this could delay the jet production. 

It is also possible instead that the jet has been present for the entire duration of the TDE.  However, due to the combination of the high density of the large cloud of circularizing TDE debris (e.g., \citealt{Bonnerot+22}) and the potential for jet precession (e.g., due to misalignment of the disk angular momentum relative to the black hole spin axis; e.g., \citealt{Stone&Loeb12}), the jet is initially choked.  At later times, as the accretion rate and gas density surrounding the black hole drop, eventually the jet is able to propagate through the debris cloud and escape.

Planned additional monitoring of \tde\ may more clearly elucidate the mechanism responsible for the delayed launch.

\section{Conclusions}
\label{sec:conc}

We presented the discovery of late- and rapidly-rising radio/mm emission from \tde\ starting at about 970 d post optical discovery, and extending to at least 1280 days. The radio emission is more luminous than previous non-relativistic TDEs, but still an order of magnitude weaker than the relativistic TDE Sw\,J1644+57. The rapid rise in luminosity coupled with the slow spectral evolution to 1250 days point to a decelerating outflow. Basic modeling assuming energy equipartition indicates that the outflow was launched $\approx 700-750$ d after optical discovery with a velocity of $\beta\approx 0.25$ (spherical outflow) or up to $\approx 0.6$ ($10^\circ$ jet). The outflow kinetic energy is at least $6\times 10^{49}$ erg.  This is the first case of a {\it delayed} mildly-relativistic outflow in a TDE, and its energy is in excess of all previous non-relativistic TDEs.  On the other hand, we find that the density of the circumnuclear environment is typical of previous TDEs, indicating that interaction with a dense medium is not the cause for the long delay.  We similarly show that an off-axis jet cannot explain the late rising radio emission.

With planned continued observations of \tde\ we will monitor the on-going evolution of the outflow and of the circumnuclear medium.
We note that the discovery of such late-time emission indicates that delayed outflows may be more common than previously expected in the TDE population.  A systematic study of a much larger sample of TDEs will be presented in Cendes et al.~in preparation.

\begin{acknowledgments}
We thank Emil Polisensky for his assistance with VLITE data, and the VLA, MeerKAT, and {\it Chandra} observatory staff for prompt responses and observations for our DDT requests.  In particular, we thank Drew Medlin for his assistance with the calibration and imaging of our wide-field VLA data.  The Berger Time-Domain Group at Harvard is supported by NSF and NASA grants. This paper makes use of the following ALMA data: ADS/JAO.ALMA\#2021.1.01210.T. TL acknowledges support from the Radboud Excellence Initiative. ALMA is a partnership of ESO (representing its member states), NSF (USA) and NINS (Japan), together with NRC (Canada), MOST and ASIAA (Taiwan), and KASI (Republic of Korea), in cooperation with the Republic of Chile. The Joint ALMA Observatory is operated by ESO, AUI/NRAO and NAOJ. The National Radio Astronomy Observatory is a facility of the National Science Foundation operated under cooperative agreement by Associated Universities, Inc. The scientific results reported in this article are based in part on observations made by the Chandra X-ray Observatory, which is operated by the Smithsonian Astrophysical Observatory for and on behalf of the National Aeronautics Space Administration under contract NAS8-03060.  The MeerKAT telescope is operated by the South African Radio Astronomy Observatory, which is a facility of the National Research Foundation, an agency of the Department of Science and Innovation.  The Australia Telescope Compact Array is part of the Australia Telescope National Facility (https://ror.org/05qajvd42) which is funded by the Australian Government for operation as a National Facility managed by CSIRO.
\end{acknowledgments}

\appendix
\label{sec:app}

As noted in \S\ref{sec:modeling}, given the unusual nature of the spectral evolution between the latest two epochs, with the peak frequency {\it increasing} by a factor of 2 in about 31 days, we consider here a model with two outflows each generating its own synchrotron emission, and peaking at $\approx 1.5$ GHz (Outflow 1) and $\approx 3$ GHz (Outflow 2).  In this scenario, Outflow 1 dominates the emission at $\approx 970-1250$ days, while Outflow 2 becomes more dominant at $\approx 1280$ days.  

To fit the data we use the same procedure described in \S\ref{sec:modeling}.  We set the MCMC priors to ensure that the time evolution of $\nu_{p,1}$ and $\nu_{p,2}$ is to either remain constant or decline, and that the peak flux density of Outflow 2 increases with time.

\begin{figure}
\begin{center}
\includegraphics[width=.45\columnwidth]{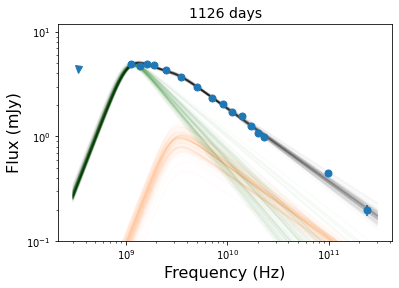}
\includegraphics[width=.45\columnwidth]{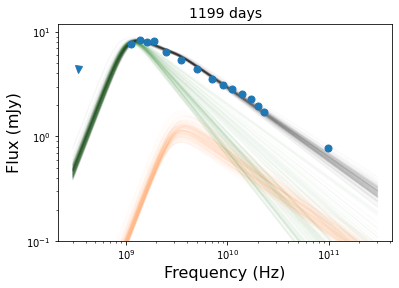}
\includegraphics[width=.45\columnwidth]{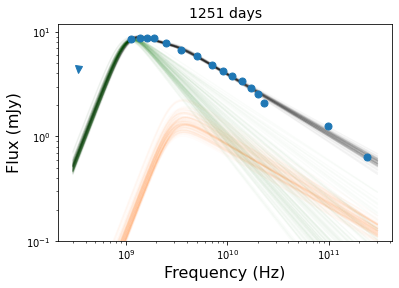}
\includegraphics[width=.45\columnwidth]{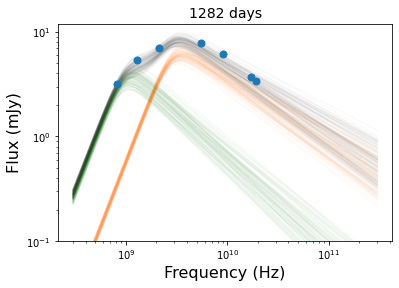}
\end{center}
\label{fig:sed}
\caption{Radio spectral energy distributions for our VLA data as in Figure \ref{fig:seds}, but for a two outflow model.  Here, Outflow 1 has a peak near $\sim1$ GHz (green lines), Outflow 2 has a peak near $\sim3$ GHz (orange lines), and they combine to give a total outflow (black lines).}
\end{figure}

We first model the first and final well sampled epochs (1126 and 1282 days) to constrain the range of $\nu_{a,1}$ and $\nu_{a,2}$.  We then model the intermediate timescales; we do not model the earliest epoch at 972 days since the SED is not well sampled enough for this two-component analysis.  The results of the models are shown in Figure~\ref{fig:sed}, and the parameters of the two outflows are provided in Table \ref{tab:sed-ouflow-1}.  

We find that Outflow 1 dominates the emission in the first three epochs, but has to rapidly fade from about 8 mJy to 3.6 mJy in the final epoch. Outflow 2 provides minimal contribution in the first three epochs, and then rises rapidly from about 0.4 mJy to 6.1 mJy in about 30 days. This would effectively correspond to Outflow 2 being generated at a later time, at about $1250-1280$ days.

Given the required rapid fading of Outflow 1, and the rapid appearance of Outflow 2, we find that while this two-outflow scenario is plausible in terms of fitting the SEDs, it does not alleviate the problem of rapid evolution on a timescale of about 30 days, hundreds of days after the appearance of radio emission.  Moreover, in this scenario the properties of Outflows 1 and 2 still require delayed, mildly-relativistic outflows.

\begin{deluxetable*}{lcccccccc}
\label{tab:sed-ouflow-1}
\tablecolumns{7}
\tablecaption{Spectral Energy Distribution Parameters for the Two-Component Model}
\tablehead{
	\colhead{$\delta t$} &
	\colhead{$F_{\nu,p,1}$} &
	\colhead{log($\nu_{p,1}$)} &
	\colhead{$p_1$} &
	\colhead{$F_{\nu,p,2}$} &
	\colhead{log($\nu_{p,2}$)} &
	\colhead{$p_2$} 	\\
    \colhead{(d)} &
    \colhead{(mJy)} &
    \colhead{(Hz)} &
    \colhead{} &
    \colhead{(mJy)} &
    \colhead{(Hz)} &
    \colhead{} 
}
\startdata
1126 & $4.84\pm 0.16$ & $9.10\pm 0.04$ & $2.7\pm 0.5$ & $0.94\pm 0.16$ & $9.55\pm 0.04$ & $2.41\pm 0.05$ \\
1199 & $7.96\pm 0.10$ & $9.07\pm 0.03$ & $2.7\pm 0.5$ & $1.17\pm 0.10$ & $9.56\pm 0.03$ & $2.35\pm 0.07$ \\
1251 & $8.00\pm 0.30$ & $9.07\pm 0.03$ & $2.7\pm 0.5$ & $1.31\pm 0.30$ & $9.55\pm 0.03$ & $2.37\pm 0.07$ \\
1282 & $3.60\pm 0.16$ & $9.05\pm 0.07$ & $2.7\pm 0.5$ & $6.06\pm 0.16$ & $9.55\pm 0.07$ & $2.30\pm 0.13$ 
\enddata
\tablecomments{$^a$ Data are combined in each observation as in Table \ref{tab:sed}.}
\end{deluxetable*}

\bibliography{sample631}{}

\begin{thebibliography}{}
\expandafter\ifx\csname natexlab\endcsname\relax\def\natexlab#1{#1}\fi
\providecommand{\url}[1]{\href{#1}{#1}}
\providecommand{\dodoi}[1]{doi:~\href{http://doi.org/#1}{\nolinkurl{#1}}}
\providecommand{\doeprint}[1]{\href{http://ascl.net/#1}{\nolinkurl{http://ascl.net/#1}}}
\providecommand{\doarXiv}[1]{\href{https://arxiv.org/abs/#1}{\nolinkurl{https://arxiv.org/abs/#1}}}

\bibitem[{{Alexander} {et~al.}(2016){Alexander}, {Berger}, {Guillochon},
  {Zauderer}, \& {Williams}}]{Alexander2016}
{Alexander}, K.~D., {Berger}, E., {Guillochon}, J., {Zauderer}, B.~A., \&
  {Williams}, P.~K.~G. 2016, \apjl, 819, L25,
  \dodoi{10.3847/2041-8205/819/2/L25}

\bibitem[{{Alexander} {et~al.}(2020){Alexander}, {van Velzen}, {Horesh}, \&
  {Zauderer}}]{Alexander2020}
{Alexander}, K.~D., {van Velzen}, S., {Horesh}, A., \& {Zauderer}, B.~A. 2020,
  arXiv e-prints, arXiv:2006.01159.
\newblock \doarXiv{2006.01159}

\bibitem[{{Alexander} {et~al.}(2017){Alexander}, {Wieringa}, {Berger},
  {Saxton}, \& {Komossa}}]{Alexander2017}
{Alexander}, K.~D., {Wieringa}, M.~H., {Berger}, E., {Saxton}, R.~D., \&
  {Komossa}, S. 2017, \apj, 837, 153, \dodoi{10.3847/1538-4357/aa6192}

\bibitem[{{Anderson} {et~al.}(2019){Anderson}, {Mooley}, {Hallinan}, {Dong},
  {Phinney}, {Horesh}, {Bourke}, {Cenko}, {Frail}, {Kulkarni}, \&
  {Myers}}]{Anderson2019}
{Anderson}, M.~M., {Mooley}, K.~P., {Hallinan}, G., {et~al.} 2019, arXiv
  e-prints, arXiv:1910.11912.
\newblock \doarXiv{1910.11912}

\bibitem[{{Baganoff} {et~al.}(2003){Baganoff}, {Maeda}, {Morris}, {Bautz},
  {Brandt}, {Cui}, {Doty}, {Feigelson}, {Garmire}, {Pravdo}, {Ricker}, \&
  {Townsley}}]{Baganoff2003}
{Baganoff}, F.~K., {Maeda}, Y., {Morris}, M., {et~al.} 2003, \apj, 591, 891,
  \dodoi{10.1086/375145}

\bibitem[{{Barbary}(2016)}]{Barbary16}
{Barbary}, K. 2016, {Extinction V0.3.0},  Zenodo, \dodoi{10.5281/zenodo.804967}

\bibitem[{{Barniol Duran} {et~al.}(2013){Barniol Duran}, {Nakar}, \&
  {Piran}}]{Duran2013}
{Barniol Duran}, R., {Nakar}, E., \& {Piran}, T. 2013, \apj, 772, 78,
  \dodoi{10.1088/0004-637X/772/1/78}

\bibitem[{{Becker}(2015)}]{Becker15}
{Becker}, A. 2015, {HOTPANTS: High Order Transform of PSF ANd Template
  Subtraction}, Astrophysics Source Code Library, record ascl:1504.004.
\newblock \doeprint{1504.004}

\bibitem[{{Berger} {et~al.}(2012){Berger}, {Zauderer}, {Pooley}, {Soderberg},
  {Sari}, {Brunthaler}, \& {Bietenholz}}]{p1}
{Berger}, E., {Zauderer}, A., {Pooley}, G.~G., {et~al.} 2012, \apj, 748, 36,
  \dodoi{10.1088/0004-637X/748/1/36}

\bibitem[{{Bonnerot} {et~al.}(2022){Bonnerot}, {Pessah}, \& {Lu}}]{Bonnerot+22}
{Bonnerot}, C., {Pessah}, M.~E., \& {Lu}, W. 2022, \apjl, 931, L6,
  \dodoi{10.3847/2041-8213/ac6950}

\bibitem[{{Cardelli} {et~al.}(1989){Cardelli}, {Clayton}, \&
  {Mathis}}]{Cardelli89}
{Cardelli}, J.~A., {Clayton}, G.~C., \& {Mathis}, J.~S. 1989, in Interstellar
  Dust, ed. L.~J. {Allamandola} \& A.~G.~G.~M. {Tielens}, Vol. 135, 5--10

\bibitem[{{Cendes} {et~al.}(2021{\natexlab{a}}){Cendes}, {Alexander}, {Berger},
  {Eftekhari}, {Williams}, \& {Chornock}}]{Cendes2021b}
{Cendes}, Y., {Alexander}, K.~D., {Berger}, E., {et~al.} 2021{\natexlab{a}},
  \apj, 919, 127, \dodoi{10.3847/1538-4357/ac110a}

\bibitem[{{Cendes} {et~al.}(2021{\natexlab{b}}){Cendes}, {Eftekhari}, {Berger},
  \& {Polisensky}}]{Cendes2021}
{Cendes}, Y., {Eftekhari}, T., {Berger}, E., \& {Polisensky}, E.
  2021{\natexlab{b}}, \apj, 908, 125, \dodoi{10.3847/1538-4357/abd323}

\bibitem[{{Clarke} {et~al.}(2016){Clarke}, {Kassim}, {Brisken}, {Helmboldt},
  {Peters}, {Ray}, {Polisensky}, \& {Giacintucci}}]{Clarke2016}
{Clarke}, T.~E., {Kassim}, N.~E., {Brisken}, W., {et~al.} 2016, in Society of
  Photo-Optical Instrumentation Engineers (SPIE) Conference Series, Vol. 9906,
  \procspie, 99065B, \dodoi{10.1117/12.2233036}

\bibitem[{{Condon} {et~al.}(1998){Condon}, {Cotton}, {Greisen}, {Yin},
  {Perley}, {Taylor}, \& {Broderick}}]{Condon1998}
{Condon}, J.~J., {Cotton}, W.~D., {Greisen}, E.~W., {et~al.} 1998, \aj, 115,
  1693, \dodoi{10.1086/300337}

\bibitem[{{De Colle} {et~al.}(2012){De Colle}, {Guillochon}, {Naiman}, \&
  {Ramirez-Ruiz}}]{DeColle2012}
{De Colle}, F., {Guillochon}, J., {Naiman}, J., \& {Ramirez-Ruiz}, E. 2012,
  \apj, 760, 103, \dodoi{10.1088/0004-637X/760/2/103}

\bibitem[{{Eftekhari} {et~al.}(2018){Eftekhari}, {Berger}, {Zauderer},
  {Margutti}, \& {Alexander}}]{Eftekhari2018}
{Eftekhari}, T., {Berger}, E., {Zauderer}, B.~A., {Margutti}, R., \&
  {Alexander}, K.~D. 2018, \apj, 854, 86, \dodoi{10.3847/1538-4357/aaa8e0}

\bibitem[{{Foreman-Mackey} {et~al.}(2013){Foreman-Mackey}, {Hogg}, {Lang}, \&
  {Goodman}}]{Foreman-Mackey2013}
{Foreman-Mackey}, D., {Hogg}, D.~W., {Lang}, D., \& {Goodman}, J. 2013, \pasp,
  125, 306, \dodoi{10.1086/670067}

\bibitem[{{Gehrels} {et~al.}(2004){Gehrels}, {Chincarini}, {Giommi}, {Mason},
  {Nousek}, {Wells}, {White}, {Barthelmy}, {Burrows}, {Cominsky}, {Hurley},
  {Marshall}, {M{\'e}sz{\'a}ros}, {Roming}, {Angelini}, {Barbier}, {Belloni},
  {Campana}, {Caraveo}, {Chester}, {Citterio}, {Cline}, {Cropper}, {Cummings},
  {Dean}, {Feigelson}, {Fenimore}, {Frail}, {Fruchter}, {Garmire}, {Gendreau},
  {Ghisellini}, {Greiner}, {Hill}, {Hunsberger}, {Krimm}, {Kulkarni}, {Kumar},
  {Lebrun}, {Lloyd-Ronning}, {Markwardt}, {Mattson}, {Mushotzky}, {Norris},
  {Osborne}, {Paczynski}, {Palmer}, {Park}, {Parsons}, {Paul}, {Rees},
  {Reynolds}, {Rhoads}, {Sasseen}, {Schaefer}, {Short}, {Smale}, {Smith},
  {Stella}, {Tagliaferri}, {Takahashi}, {Tashiro}, {Townsley}, {Tueller},
  {Turner}, {Vietri}, {Voges}, {Ward}, {Willingale}, {Zerbi}, \&
  {Zhang}}]{Gehrels04}
{Gehrels}, N., {Chincarini}, G., {Giommi}, P., {et~al.} 2004, \apj, 611, 1005,
  \dodoi{10.1086/422091}

\bibitem[{{Generozov} {et~al.}(2017){Generozov}, {Mimica}, {Metzger}, {Stone},
  {Giannios}, \& {Aloy}}]{Generozov2017}
{Generozov}, A., {Mimica}, P., {Metzger}, B.~D., {et~al.} 2017, \mnras, 464,
  2481, \dodoi{10.1093/mnras/stw2439}

\bibitem[{{Giannios} \& {Metzger}(2011)}]{Giannios11}
{Giannios}, D., \& {Metzger}, B.~D. 2011, \mnras, 416, 2102,
  \dodoi{10.1111/j.1365-2966.2011.19188.x}

\bibitem[{{Gillessen} {et~al.}(2019){Gillessen}, {Plewa}, {Widmann}, {von
  Fellenberg}, {Schartmann}, {Habibi}, {Jimenez Rosales}, {Baub{\"o}ck},
  {Dexter}, {Gao}, {Waisberg}, {Eisenhauer}, {Pfuhl}, {Ott}, {Burkert}, {de
  Zeeuw}, \& {Genzel}}]{Gillessen2019}
{Gillessen}, S., {Plewa}, P.~M., {Widmann}, F., {et~al.} 2019, \apj, 871, 126,
  \dodoi{10.3847/1538-4357/aaf4f8}

\bibitem[{{Gomez} {et~al.}(2020){Gomez}, {Nicholl}, {Short}, {Margutti},
  {Alexander}, {Blanchard}, {Berger}, {Eftekhari}, {Schulze}, {Anderson},
  {Arcavi}, {Chornock}, {Cowperthwaite}, {Galbany}, {Herzog}, {Hiramatsu},
  {Hosseinzadeh}, {Laskar}, {M{\"u}ller Bravo}, {Patton}, \&
  {Terreran}}]{Gomez2020}
{Gomez}, S., {Nicholl}, M., {Short}, P., {et~al.} 2020, \mnras, 497, 1925,
  \dodoi{10.1093/mnras/staa2099}

\bibitem[{{Goodwin} {et~al.}(2022){Goodwin}, {van Velzen}, {Miller-Jones},
  {Mummery}, {Bietenholz}, {Wederfoort}, {Hammerstein}, {Bonnerot}, {Hoffmann},
  \& {Yan}}]{Goodwin2022}
{Goodwin}, A.~J., {van Velzen}, S., {Miller-Jones}, J.~C.~A., {et~al.} 2022,
  \mnras, \dodoi{10.1093/mnras/stac333}

\bibitem[{{Granot} \& {Sari}(2002)}]{Granot2002}
{Granot}, J., \& {Sari}, R. 2002, \apj, 568, 820, \dodoi{10.1086/338966}

\bibitem[{{Guillochon} {et~al.}(2016){Guillochon}, {McCourt}, {Chen},
  {Johnson}, \& {Berger}}]{Guillochon2016}
{Guillochon}, J., {McCourt}, M., {Chen}, X., {Johnson}, M.~D., \& {Berger}, E.
  2016, \apj, 822, 48, \dodoi{10.3847/0004-637X/822/1/48}

\bibitem[{{Guillochon} \& {Ramirez-Ruiz}(2013)}]{Guillochon2013}
{Guillochon}, J., \& {Ramirez-Ruiz}, E. 2013, \apj, 767, 25,
  \dodoi{10.1088/0004-637X/767/1/25}

\bibitem[{{Hammerstein} {et~al.}(2022){Hammerstein}, {van Velzen}, {Gezari},
  {Cenko}, {Yao}, {Ward}, {Frederick}, {Villanueva}, {Somalwar}, {Graham},
  {Kulkarni}, {Stern}, {Bellm}, {Dekany}, {Drake}, {Groom}, {Kasliwal}, {Kool},
  {Masci}, {Medford}, \& {van Roestel}}]{Hammerstein2022}
{Hammerstein}, E., {van Velzen}, S., {Gezari}, S., {et~al.} 2022, arXiv
  e-prints, arXiv:2203.01461.
\newblock \doarXiv{2203.01461}

\bibitem[{{Heasarc}(2014)}]{Nasa14}
{Heasarc}, N. H. E. A. S. A. R.~C. 2014, {HEAsoft: Unified Release of FTOOLS
  and XANADU}.
\newblock \doeprint{1408.004}

\bibitem[{{Horesh} {et~al.}(2021{\natexlab{a}}){Horesh}, {Cenko}, \&
  {Arcavi}}]{Horesh2021}
{Horesh}, A., {Cenko}, S.~B., \& {Arcavi}, I. 2021{\natexlab{a}}, Nature
  Astronomy, 5, 491, \dodoi{10.1038/s41550-021-01300-8}

\bibitem[{{Horesh} {et~al.}(2021{\natexlab{b}}){Horesh}, {Sfaradi}, {Fender},
  {Green}, {Williams}, \& {Bright}}]{Horesh2021b}
{Horesh}, A., {Sfaradi}, I., {Fender}, R., {et~al.} 2021{\natexlab{b}}, \apjl,
  920, L5, \dodoi{10.3847/2041-8213/ac25fe}

\bibitem[{{Horesh} {et~al.}(2018){Horesh}, {Sfaradi}, {Bright}, {Williams},
  {Fender}, {Titterington}, {Green}, {Perrott}, {Van Velzen}, \&
  {Gezari}}]{Horesh2018}
{Horesh}, A., {Sfaradi}, I., {Bright}, J., {et~al.} 2018, The Astronomer's
  Telegram, 12271, 1

\bibitem[{{Horesh} {et~al.}(2022){Horesh}, {Burger}, {Sfaradi}, {Gaensler},
  {Murphy}, {Leung}, {Obrien}, {Kaplan}, \& {VAST Collaboration}}]{Horesh2022}
{Horesh}, A., {Burger}, N., {Sfaradi}, I., {et~al.} 2022, The Astronomer's
  Telegram, 15307, 1

\bibitem[{{Kalberla} {et~al.}(2005){Kalberla}, {Burton}, {Hartmann}, {Arnal},
  {Bajaja}, {Morras}, \& {P{\"o}ppel}}]{Kalberla05}
{Kalberla}, P.~M.~W., {Burton}, W.~B., {Hartmann}, D., {et~al.} 2005, \aap,
  440, 775, \dodoi{10.1051/0004-6361:20041864}

\bibitem[{{Kelley} {et~al.}(2014){Kelley}, {Tchekhovskoy}, \&
  {Narayan}}]{Kelley2014}
{Kelley}, L.~Z., {Tchekhovskoy}, A., \& {Narayan}, R. 2014, \mnras, 445, 3919,
  \dodoi{10.1093/mnras/stu2041}

\bibitem[{{Komossa}(2015)}]{Komossa2015}
{Komossa}, S. 2015, Journal of High Energy Astrophysics, 7, 148,
  \dodoi{10.1016/j.jheap.2015.04.006}

\bibitem[{{Lacy} {et~al.}(2020){Lacy}, {Baum}, {Chandler}, {Chatterjee},
  {Clarke}, {Deustua}, {English}, {Farnes}, {Gaensler}, {Gugliucci},
  {Hallinan}, {Kent}, {Kimball}, {Law}, {Lazio}, {Marvil}, {Mao}, {Medlin},
  {Mooley}, {Murphy}, {Myers}, {Osten}, {Richards}, {Rosolowsky}, {Rudnick},
  {Schinzel}, {Sivakoff}, {Sjouwerman}, {Taylor}, {White}, {Wrobel},
  {Andernach}, {Beasley}, {Berger}, {Bhatnager}, {Birkinshaw}, {Bower},
  {Brandt}, {Brown}, {Burke-Spolaor}, {Butler}, {Comerford}, {Demorest}, {Fu},
  {Giacintucci}, {Golap}, {G{\"u}th}, {Hales}, {Hiriart}, {Hodge}, {Horesh},
  {Ivezi{\'c}}, {Jarvis}, {Kamble}, {Kassim}, {Liu}, {Loinard}, {Lyons},
  {Masters}, {Mezcua}, {Moellenbrock}, {Mroczkowski}, {Nyland},
  {O{\textquoteright}Dea}, {O{\textquoteright}Sullivan}, {Peters}, {Radford},
  {Rao}, {Robnett}, {Salcido}, {Shen}, {Sobotka}, {Witz}, {Vaccari}, {van
  Weeren}, {Vargas}, {Williams}, \& {Yoon}}]{Lacy2020}
{Lacy}, M., {Baum}, S.~A., {Chandler}, C.~J., {et~al.} 2020, \pasp, 132,
  035001, \dodoi{10.1088/1538-3873/ab63eb}

\bibitem[{{Liska} {et~al.}(2020){Liska}, {Tchekhovskoy}, \&
  {Quataert}}]{Liska2020}
{Liska}, M., {Tchekhovskoy}, A., \& {Quataert}, E. 2020, \mnras, 494, 3656,
  \dodoi{10.1093/mnras/staa955}

\bibitem[{{McMullin} {et~al.}(2007){McMullin}, {Waters}, {Schiebel}, {Young},
  \& {Golap}}]{McMullin2007}
{McMullin}, J.~P., {Waters}, B., {Schiebel}, D., {Young}, W., \& {Golap}, K.
  2007, in Astronomical Society of the Pacific Conference Series, Vol. 376,
  Astronomical Data Analysis Software and Systems XVI, ed. R.~A. {Shaw},
  F.~{Hill}, \& D.~J. {Bell}, 127

\bibitem[{{Metzger} {et~al.}(2012){Metzger}, {Giannios}, \&
  {Mimica}}]{Metzger2012}
{Metzger}, B.~D., {Giannios}, D., \& {Mimica}, P. 2012, \mnras, 420, 3528,
  \dodoi{10.1111/j.1365-2966.2011.20273.x}

\bibitem[{{Mimica} {et~al.}(2015){Mimica}, {Giannios}, {Metzger}, \&
  {Aloy}}]{Mimica2015}
{Mimica}, P., {Giannios}, D., {Metzger}, B.~D., \& {Aloy}, M.~A. 2015, \mnras,
  450, 2824, \dodoi{10.1093/mnras/stv825}

\bibitem[{{Murphy} {et~al.}(2021){Murphy}, {Kaplan}, {Stewart}, {O'Brien},
  {Lenc}, {Pintaldi}, {Pritchard}, {Dobie}, {Fox}, {Leung}, {An}, {Bell},
  {Broderick}, {Chatterjee}, {Dai}, {d'Antonio}, {Doyle}, {Gaensler}, {Heald},
  {Horesh}, {Jones}, {McConnell}, {Moss}, {Raja}, {Ramsay}, {Ryder}, {Sadler},
  {Sivakoff}, {Wang}, {Wang}, {Wheatland}, {Whiting}, {Allison}, {Anderson},
  {Ball}, {Bannister}, {Bock}, {Bolton}, {Bunton}, {Chekkala}, {Chippendale},
  {Cooray}, {Gupta}, {Hayman}, {Jeganathan}, {Koribalski}, {Lee-Waddell},
  {Mahony}, {Marvil}, {McClure-Griffiths}, {Mirtschin}, {Ng}, {Pearce},
  {Phillips}, \& {Voronkov}}]{Murphy2021}
{Murphy}, T., {Kaplan}, D.~L., {Stewart}, A.~J., {et~al.} 2021, \pasa, 38,
  e054, \dodoi{10.1017/pasa.2021.44}

\bibitem[{{Nakar} \& {Granot}(2007)}]{NakarGranot2007}
{Nakar}, E., \& {Granot}, J. 2007, \mnras, 380, 1744,
  \dodoi{10.1111/j.1365-2966.2007.12245.x}

\bibitem[{{Nakar} \& {Piran}(2011)}]{Nakar2011}
{Nakar}, E., \& {Piran}, T. 2011, \nat, 478, 82, \dodoi{10.1038/nature10365}

\bibitem[{{Rees}(1988)}]{Rees1988}
{Rees}, M.~J. 1988, \nat, 333, 523, \dodoi{10.1038/333523a0}

\bibitem[{{Roming} {et~al.}(2005){Roming}, {Kennedy}, {Mason}, {Nousek}, {Ahr},
  {Bingham}, {Broos}, {Carter}, {Hancock}, {Huckle}, {Hunsberger}, {Kawakami},
  {Killough}, {Koch}, {McLelland}, {Smith}, {Smith}, {Soto}, {Boyd},
  {Breeveld}, {Holland}, {Ivanushkina}, {Pryzby}, {Still}, \&
  {Stock}}]{Roming05}
{Roming}, P. W.~A., {Kennedy}, T.~E., {Mason}, K.~O., {et~al.} 2005, \ssr, 120,
  95, \dodoi{10.1007/s11214-005-5095-4}

\bibitem[{{Russell} {et~al.}(2015){Russell}, {Fabian}, {McNamara}, \&
  {Broderick}}]{Russell2015}
{Russell}, H.~R., {Fabian}, A.~C., {McNamara}, B.~R., \& {Broderick}, A.~E.
  2015, \mnras, 451, 588, \dodoi{10.1093/mnras/stv954}

\bibitem[{{Sari} {et~al.}(1998){Sari}, {Piran}, \& {Narayan}}]{Sari1998}
{Sari}, R., {Piran}, T., \& {Narayan}, R. 1998, \apjl, 497, L17,
  \dodoi{10.1086/311269}

\bibitem[{{Short} {et~al.}(2020){Short}, {Nicholl}, {Lawrence}, {Gomez},
  {Arcavi}, {Wevers}, {Leloudas}, {Schulze}, {Anderson}, {Berger}, {Blanchard},
  {Burke}, {Segura}, {Charalampopoulos}, {Chornock}, {Galbany}, {Gromadzki},
  {Herzog}, {Hiramatsu}, {Horne}, {Hosseinzadeh}, {Howell}, {Ihanec},
  {Inserra}, {Kankare}, {Maguire}, {McCully}, {M{\"u}ller Bravo}, {Onori},
  {Sollerman}, \& {Young}}]{Short2020}
{Short}, P., {Nicholl}, M., {Lawrence}, A., {et~al.} 2020, \mnras, 498, 4119,
  \dodoi{10.1093/mnras/staa2065}

\bibitem[{{Stein} {et~al.}(2021){Stein}, {Velzen}, {Kowalski}, {Franckowiak},
  {Gezari}, {Miller-Jones}, {Frederick}, {Sfaradi}, {Bietenholz}, {Horesh},
  {Fender}, {Garrappa}, {Ahumada}, {Andreoni}, {Belicki}, {Bellm},
  {B{\"o}ttcher}, {Brinnel}, {Burruss}, {Cenko}, {Coughlin}, {Cunningham},
  {Drake}, {Farrar}, {Feeney}, {Foley}, {Gal-Yam}, {Golkhou}, {Goobar},
  {Graham}, {Hammerstein}, {Helou}, {Hung}, {Kasliwal}, {Kilpatrick}, {Kong},
  {Kupfer}, {Laher}, {Mahabal}, {Masci}, {Necker}, {Nordin}, {Perley},
  {Rigault}, {Reusch}, {Rodriguez}, {Rojas-Bravo}, {Rusholme}, {Shupe},
  {Singer}, {Sollerman}, {Soumagnac}, {Stern}, {Taggart}, {van Santen}, {Ward},
  {Woudt}, \& {Yao}}]{Stein2020}
{Stein}, R., {Velzen}, S.~v., {Kowalski}, M., {et~al.} 2021, Nature Astronomy,
  \dodoi{10.1038/s41550-020-01295-8}

\bibitem[{{Stone} \& {Loeb}(2012)}]{Stone&Loeb12}
{Stone}, N., \& {Loeb}, A. 2012, \prl, 108, 061302,
  \dodoi{10.1103/PhysRevLett.108.061302}

\bibitem[{{Stone} {et~al.}(2013){Stone}, {Sari}, \& {Loeb}}]{Stone2013}
{Stone}, N., {Sari}, R., \& {Loeb}, A. 2013, \mnras, 435, 1809,
  \dodoi{10.1093/mnras/stt1270}

\bibitem[{{Tchekhovskoy} {et~al.}(2014){Tchekhovskoy}, {Metzger}, {Giannios},
  \& {Kelley}}]{Tchekhovskoy2014}
{Tchekhovskoy}, A., {Metzger}, B.~D., {Giannios}, D., \& {Kelley}, L.~Z. 2014,
  \mnras, 437, 2744, \dodoi{10.1093/mnras/stt2085}

\bibitem[{{Williams} {et~al.}(2017){Williams}, {Gizis}, \&
  {Berger}}]{Williams2017}
{Williams}, P.~K.~G., {Gizis}, J.~E., \& {Berger}, E. 2017, \apj, 834, 117,
  \dodoi{10.3847/1538-4357/834/2/117}

\bibitem[{{Yalinewich} {et~al.}(2019){Yalinewich}, {Steinberg}, {Piran}, \&
  {Krolik}}]{Yalinewich2019}
{Yalinewich}, A., {Steinberg}, E., {Piran}, T., \& {Krolik}, J.~H. 2019,
  \mnras, 487, 4083, \dodoi{10.1093/mnras/stz1567}

\bibitem[{{Zauderer} {et~al.}(2013){Zauderer}, {Berger}, {Margutti}, {Pooley},
  {Sari}, {Soderberg}, {Brunthaler}, \& {Bietenholz}}]{p2}
{Zauderer}, B.~A., {Berger}, E., {Margutti}, R., {et~al.} 2013, \apj, 767, 152,
  \dodoi{10.1088/0004-637X/767/2/152}

\bibitem[{{Zauderer} {et~al.}(2011){Zauderer}, {Berger}, {Soderberg}, {Loeb},
  {Narayan}, {Frail}, {Petitpas}, {Brunthaler}, {Chornock}, {Carpenter},
  {Pooley}, {Mooley}, {Kulkarni}, {Margutti}, {Fox}, {Nakar}, {Patel},
  {Volgenau}, {Culverhouse}, {Bietenholz}, {Rupen}, {Max-Moerbeck}, {Readhead},
  {Richards}, {Shepherd}, {Storm}, \& {Hull}}]{Zauderer2011}
{Zauderer}, B.~A., {Berger}, E., {Soderberg}, A.~M., {et~al.} 2011, \nat, 476,
  425, \dodoi{10.1038/nature10366}

\end{thebibliography}
\bibliographystyle{aasjournal}

\end{document}